\documentclass[12pt,journal,draftclsnofoot,onecolumn,english]{IEEEtran}
\usepackage{amsmath,amsfonts}
\usepackage{amssymb}
\usepackage{amsthm}
\usepackage{makecell}
\usepackage{tabularx}
\usepackage{algorithmic}
\usepackage{array}
\usepackage[caption=false,font=normalsize,labelfont=sf,textfont=sf]{subfig}
\usepackage{graphicx}
\usepackage[justification=centering]{caption}
\usepackage{mathrsfs}
\usepackage{textcomp}
\usepackage{stfloats}
\usepackage{url}
\usepackage{bm}
\usepackage{cite}
\usepackage{color,xcolor}
\usepackage{verbatim}
\usepackage{graphicx}
\usepackage{pstricks}
\usepackage{epstopdf}
\usepackage{bm,bbm}
\newtheorem{theorem}{Theorem}
\newtheorem{lemma}{Lemma}

\newtheorem{corollary}{Corollary}

\newtheorem{remark}{Remark}

\def\BibTeX{{\rm B\kern-.05em{\sc i\kern-.025em b}\kern-.08em
		T\kern-.1667em\lower.7ex\hbox{E}\kern-.125emX}}
\usepackage{balance}
\begin{document}
	\title{Information-Theoretic Limits of Bistatic Integrated Sensing and Communication}
		\author{Tian~Jiao,  Kai~Wan,~\IEEEmembership{Member,~IEEE}, Zhiqiang~Wei,~\IEEEmembership{Member,~IEEE}, Yanlin~Geng,~\IEEEmembership{Member,~IEEE}, Yonglong~Li,~\IEEEmembership{Member,~IEEE}, Zai~Yang,~\IEEEmembership{Senior~Member,~IEEE}, and		    Giuseppe~Caire,~\IEEEmembership{Fellow,~IEEE} 
			\thanks{This work was supported in part by the major key project of Peng Cheng Laboratory under grant PCL2023AS1-2, in part by the National Natural Science Foundation of China (NSFC-12141107), in part by the National Key R\&D Program of China (Grant No. 2021YFA1000500), in part by Huawei Tech. Co., Ltd, and in part by the European Research Council under the ERC Advanced Grant N.789190, CARENET.  A preliminary version of this work was presented in part at the IEEE Information Theory Workshop (ITW), 2024 \nocite{jiao2024rate}.
				\emph{(Corresponding author: Zai~Yang.)}}
		    \thanks{T.~Jiao, Z. Wei, Y. Li, and Z. Yang are with the School of Mathematics and Statistics, Xi'an Jiaotong University, Xi'an 710049, China. 
		    Z. Wei and Z. Yang are also with the Peng Cheng Laboratory, Shenzhen, Guangdong 518055, China, and also with the Pazhou Laboratory (Huangpu), Guangzhou, Guangdong 510555, China (e-mail: tianjiao@stu.xjtu.edu.cn, zhiqiang.wei@xjtu.edu.cn, liyonglong@xjtu.edu.cn, yangzai@xjtu.edu.cn). 
		   }
			 \thanks{K.~Wan is with the School of Electronic Information and Communications, Huazhong University of Science and Technology,  Wuhan 430074, China (email:kai\_wan@hust.edu.cn). 
			 	}
		    \thanks{Y.~Geng is with the State Key Laboratory of ISN, Xidian University, Xi'an 710049, China (e-mail: ylgeng@xidian.edu.cn). 
}  
		    \thanks{G. Caire is with the Electrical Engineering and Computer Science Department, Technische Universität Berlin, 10587 Berlin, Germany (e-mail: caire@tu-berlin.de). 
	    }}

	\maketitle
	
	\begin{abstract}
		Bistatic sensing refers to scenarios where the transmitter (illuminating the target) and the sensing receiver (estimating the target state) are physically separated, in contrast to monostatic sensing, where both functions are co-located. In practical settings, bistatic sensing may be required either due to inherent system constraints or as a means to mitigate the strong self-interference encountered in monostatic configurations. 
			A key practical challenge in bistatic radio-frequency radar systems is the synchronization and calibration of the separate transmitter and sensing receiver.
			In this paper, we are not concerned with these signal processing aspects and take a complementary information-theoretic perspective on bistatic integrated sensing and communication (ISAC). Namely, we aim to characterize the capacity-distortion function—the fundamental tradeoff between communication capacity and sensing accuracy. We consider a general discrete channel model for a bistatic ISAC system and derive a multi-letter representation of its capacity-distortion function. Then, we establish single-letter upper and lower bounds and provide exact single-letter characterizations for degraded bistatic ISAC channels. Furthermore, we extend our analysis to a bistatic ISAC broadcast channel and derive the capacity-distortion region with a single-letter characterization in the degraded case. Numerical examples illustrate the theoretical results, highlighting the benefits of ISAC over separate communication and sensing, as well as the role of leveraging communication to assist sensing in bistatic systems.			

	\end{abstract}
	

	\section{Introduction}
	\IEEEPARstart{I}{ntegrated} sensing and communication (ISAC) has emerged as a key technology and research area in future-generation wireless networks (beyond 5G, 6G) since many practical scenarios place high demands on sensing and communication \cite{liu2020joint,liu2022survey}. For example, self-driving technologies not only require a high data rate for obtaining important information such as media messages, ultra-high-resolution maps, and real-time traffic information, but also need sensing functionality to provide robust and high-resolution obstacle detection \cite{bilik2019rise}. Besides, following the trend of wireless technology, 
	with the use of increasingly larger signal bandwidths~\cite{xiao2017millimeter,zhao2019mid} and antenna arrays~\cite{Marzetta2010massiveMIMO,gao2013massive,lu2014overview,larsson2014massive}, 			
	communication signals in future systems will be able to provide  high resolution both in the delay (i.e., range) and angle domains, and hence can be used  for high-precision sensing. ISAC in general refers to an ensemble of approaches to integrate the sensing and communication functionalities on a single platform such that they can share the same transmission resource (time slots, bandwidth, and power) and the same hardware compared to separate solutions \cite{liu2020joint,liu2022survey,liu2022integrated}. 
	
 In parallel to the ISAC research centered on wireless communication applications \cite{sturm2011waveform,xiao2022waveform,gaudio2020effectiveness,dokhanchi2019mmwave,gao2022integrated,elbir2022rise,sankar2022beamforming, dehkordi2023beam, dehkordi2024multistatic}, the topic has also been the focus of recent information-theoretic research. The authors in \cite{ahmadipour2022information,kobayashi2018joint} studied the monostatic ISAC model from an information-theoretic perspective and characterize the optimal tradeoff between  the capacity of reliable communication and the distortion of state estimation. In the considered system, a transmitter aims to transmit messages to a receiver through a memoryless state-dependent channel where the state sequence is independent and identically distributed (i.i.d.); meanwhile, the transmitter also aims to estimate the state sequence of the receiver through the backscattered signal, which is modeled as a causal generalized feedback. 
	Following the seminal work in~\cite{kobayashi2018joint,ahmadipour2022information}, various information-theoretic ISAC works were proposed, which could be generally classified into two frameworks according to the considered channel models: 
	\begin{itemize}
		\item In the first framework, as in the original work~\cite{kobayashi2018joint,ahmadipour2022information}, 
		the state is i.i.d. over time slots (i.e.,  the state-dependent channel is memoryless) or i.i.d. across blocks of length $T$ (i.e., the state varies after each $T$ time slots, called  block-fading state, and thus the state-dependent channel has 
		``in-block memory'' \cite{kramer2014information}). 
		In this framework, it is suitable to model state-dependent channels where the state evolves according to some stationary ergodic stochastic process and average estimation distortion is meaningful.
		Specifically, the authors in \cite{li2023capacity} extended the monostatic ISAC model with memoryless channel  to channel with a new state-dependent in-block memory channel  referred to as the binary beam-pointing (BBP) channel, and derived the corresponding capacity-distortion tradeoff.
		Vector Gaussian channel with in-block memory was considered in~\cite{xiong2023fundamental,liu2023deterministic}, where two fundamental tradeoffs between sensing and communication,  subspace tradeoff and the random-deterministic tradeoff,  were identified.  
		The capacity-distortion region of monostatic ISAC when the receiver have imperfect state knowledge is  studied in \cite{liu2022information}, where the sensing parameters and channel state are not completely consistent. The papers    \cite{kobayashi2019joint,ahmadipour2022coding,liu2022generalized} contributed in the capacity-distortion tradeoff for ISAC in the multiple access channel (MAC), by proposing different cooperative communication and cooperative sensing methods respectively, thus giving the corresponding achievable rate-distortion regions.
		\item In the second framework, as considered in~\cite{joudeh2022joint,wu2022joint,chang2023rate},   the state  is a discrete deterministic unknown and remains constant over the whole transmission block length, and the corresponding tradeoff between communication and sensing is expressed in terms of the communication rate versus state detection-error exponent. In this framework, since the state is constant, it is suitable to model ''compound channel'' scenarios, where the channel transition probability assignment can be one element in a discrete set of possible probabilities, e.g., modeling the presence or absence of a target. 
	\end{itemize}
	
	In this paper,  
	we focus on the bistatic sensing system, where the state is i.i.d. over time slots. 
	According to the distribution of transceivers, sensing systems can be classified as monostatic, bistatic, and multistatic. In particular, a sensing system with physically-colocated transmit and receiver antennas is called a monostatic sensing system, and in many cases the same antenna array is used for both transmitting and receiving. A sensing system with physically-separated transmitter and receiver antennas is called a bistatic sensing system. If multiple separated receivers are employed with one transmitter, the sensing system is called multistatic \cite{skolnik1961analysis}. 
	 Although the bistatic sensing system is generally more complicated to implement than the monostatic sensing system  \cite{willis2005bistatic}, the advantages of the bistatic sensing system in suppressing self-interference and enhancing target detection and localization accuracy have stimulated continued research. For example, a target that is designed to minimize backscatter by reflecting radar energy in other directions is difficult to be detected by a monostatic sensing system, while it may be easily detected by a bistatic sensing system \cite{burkholder2003comparison}. Also, interference from the transmit antenna to the sensing receiver antenna may be significant in the monostatic sensing system, while it is negligible in the bistatic sensing system as the transmitter and sensing receiver are far apart. A crucial problem associated with bistatic sensing systems is the synchronization issue, which requires precise alignment of time and frequency between the physically-separated transmitter and receiver. In such a system, the transmitter and receiver have independent positions and motion states, which may lead to synchronization issues in multiple dimensions such as time, frequency, and phase \cite{weib2004synchronisation}. Common methods used to address synchronization challenges include global positioning system (GPS) clock synchronization \cite{weib2004synchronisation,yulin2008precise},  phase-locked loops (PLLs) phase synchronization\cite{stensby1997phase, park2024lfm}, and adaptive synchronization \cite{wang2007approach}, etc. 
	
	Motivated by the appealing benefits of bistatic radar, in this work, we study the bistatic ISAC system from an information-theoretic perspective, along the lines of previous information-theoretic ISAC works reviewed before. 
		\begin{figure} 
		\centering
		\includegraphics[width=2.5in]{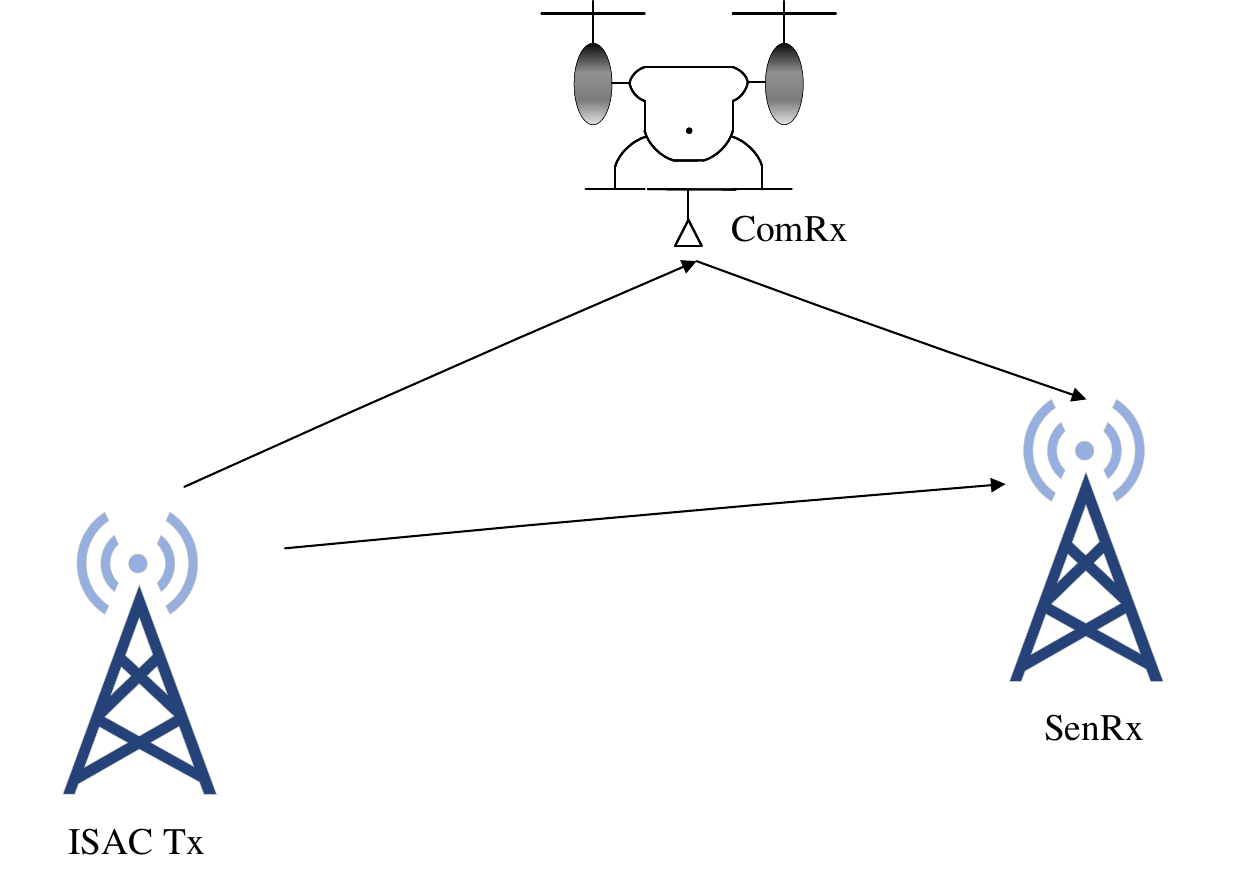}
		\caption{The bistatic ISAC system model.} 
		\label{1}
	\end{figure}
 As seen in Fig. \ref{1}, the considered bistatic ISAC system consists of a transmitter (ISAC Tx), a communication receiver (ComRx), and a sensing receiver (SenRx). The ISAC Tx sends a codeword to convey some information to the ComRx who knows the channel states perfectly. At the same time, the SenRx at another location receives both the radiated signals of the ISAC Tx and the reflected signals from the ComRx to perform the sensing task to estimate the channel states.  
 	For this scenario, we consider a general discrete channel model.
 	As such, synchronization issues are irrelevant to the model we are concerned with. In practice, these issues can be assumed to be solved by the aforementioned existing techniques.

	 It is worth pointing out that the main conceptual difference between the bistatic ISAC information-theoretic problem considered here and the existing bistatic model considered in \cite{ahmadipour2023strong} is that in our case, the SenRx is aware of the communication codebook, but it is unaware of the sent message (i.e., which codeword is sent). 
	 Therefore, it is expected  that in a bistatic ISAC system, the randomness of the communication signal exerts a greater impact on sensing than that in a monostatic ISAC system, implying a different communication-sensing performance tradeoff.
	
	Note that the paper \cite{chen2025fundamental} followed our preprint and studied a similar ISAC problem. Specifically, the authors of \cite{chen2025fundamental} selected logarithmic loss to measure the quality of a soft estimate and derived the corresponding capacity-distortion function of the ISAC model. In contrast, we study the capacity-distortion function based on a more general distortion metric, including part of the results in \cite{chen2025fundamental} as a special case of our results.
	
	  Under the information-theoretic framework, our main contributions are as follows.
	\begin{itemize}
		\item A multi-letter representation of the capacity-distortion function of a bistatic ISAC system is presented and then we transform this infinite sequence of finite-dimensional optimization problems into an infinite-dimensional optimization problem.
		
		\item We propose a single-letter lower bound based on the superposition coding scheme and partial-decoding-based estimation strategy and a single-letter upper bound by introducing a genie-aided estimator.  In particular, for the special case where the channel to SenRx is  degraded with respect to the channel to ComRx, we provide the single-letter characterization of the capacity-distortion function by showing that the derived upper and lower bounds coincide.

		
		\item The results are extended to a bistatic ISAC broadcast channel model and the capacity-distortion region with a single-letter characterization is obtained for the degraded case.
		
		\item  Examples are provided to explicitly demonstrate the significance of our proposed bounds and illustrate the benefits of exploiting communications to assist sensing for a bistatic system. In particular, for the channel model in Example 1, we present the capacity-distortion function in a simplified form.
	\end{itemize} 

	
	\paragraph*{Paper organization}
	The rest of the paper is organized as follows. Section \ref{secmo} introduces the system model and defines the capacity-distortion function. Section \ref{secp} gives a multi-letter representation of capacity-distortion function, derives some single-letter lower and upper bounds on the capacity-distortion function and the single-letter characterization of the capacity-distortion function in the degraded case, and extends the results to a  bistatic ISAC broadcast channel model. Section \ref{secex} gives the two examples to show explicitly the theoretical results on the capacity-distortion function. Section \ref{secco} concludes this paper.
	
	\paragraph*{Notation convention}
	Upper-case letters represent random variables, and lower-case letters represent their realizations. Calligraphic letters denote sets, e.g., $\mathcal{X}$.
	$\mathbb{R}_{+}$ and $\mathbb{N}_+$ represent the sets of non-negative real numbers and non-negative integer, respectively. $X^n$ denotes the tuple of random variables $(X_1,X_2,\cdots,X_n)$, $\mathbb{E}[X]$ denotes the expectation for a random variable $X$, and $\oplus$ denotes the modulo-2 sum.
	
	\section{System Model}\label{secmo}

	In this section, we introduce the bistatic ISAC system model, which is modeled as a state-dependent memoryless channel (SDMC) with two receivers.  The SDMC model with two receivers consists of four finite sets $\mathcal{X}, \mathcal{S}, \mathcal{Y}, \mathcal{Z}$  and a collection of conditional probability mass functions (pmf) $p\left(y,z|x,s\right)$ over $\mathcal{Y} \times \mathcal{Z}$. The state sequence $S^n=(S_1,\ldots,S_n)$ is i.i.d. according to a given state distribution $P_S(\cdot)$. We assume $S^n$ is perfectly and noncausally available to ComRx but unknown to SenRx.
	\begin{figure}[h]
		\centering
		\includegraphics[width=3.5in]{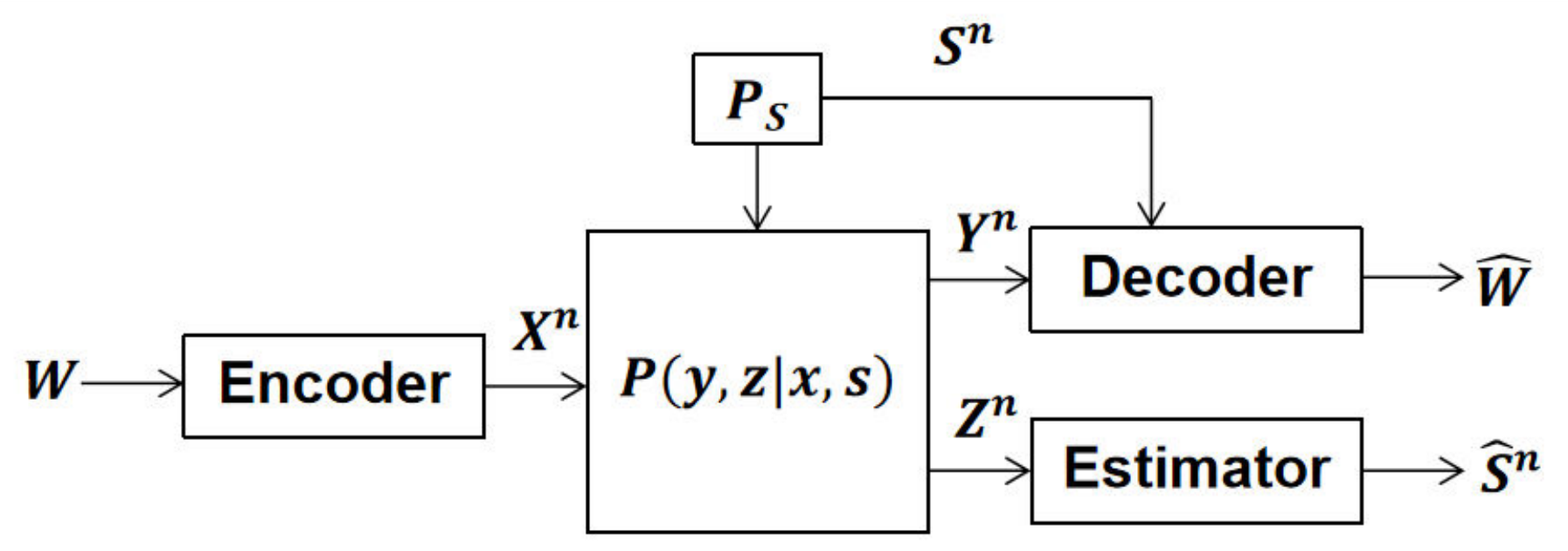}
		\caption{Two-receiver SDMC model.}
		\label{MO1}
	\end{figure}
 
 As shown in Fig.\ref{MO1}, the transmitter sends $X^n$ through the SDMC with two receivers. After receiving $Y^n$, combing with $S^n$ the ComRx finds the estimate of the message. After receiving $Z^n$, the SenRx estimates the states of the SDMC with two receivers. 
	
	A $\left(2^{n \mathrm{R}}, n\right)$ code for the above SDMC with two receivers  consists of
	
	1) a message set $\left[1: 2^{n R}\right]$;
	
	2) an encoder that assigns a codeword $x^n(w)$ to each message $w \in\left[1: 2^{n R}\right]$;
	
	3) a decoder assigns an estimate $\hat{w} \in\left[1: 2^{n R}\right]$ to each received sequence $y^n$;
	
	4) a state estimator $h: \mathcal{Z}^n \rightarrow \mathcal{S}^n $ that outputs $\hat{s}^n=h(z^n )$ as the estimation of the state sequence $s^n$ to each received sequence $z^n$.

	We assume that the message $W$ is uniformly distributed over $\left[1: 2^{n R}\right]$. The performance of the decoder is measured by the average probability of error $P_{e}^{(n)}=\mathrm{Pr}\big\{\hat{W} \neq W\big\}$.
	The accuracy of the state estimation is measured by the average expected  distortion\footnote{Since the SenRx is unaware of the sent message, we define the distortion at the SenRx by adopting the average expected distortion rather than the maximum distortion relying on the sent message, where the maximum expected distortion is defined in Remark \ref{maxdis}.}
	\begin{align}\label{D}
		D^{(n)}:=\frac{1}{n}\mathbb{E}\big[d(S^n, \hat{S}^n)\big]=\frac{1}{n} \sum_{i=1}^n \mathbb{E}\big[d(S_i, \hat{S}_i)\big],
	\end{align}
	where $d: \mathcal{S} \times \hat{\mathcal{S}} \mapsto \mathbb{R}^{+}$ is a bounded distortion function with
	\begin{align}
		d_{\max}= \max _{(s, \hat{s}) \in \mathcal{S} \times \hat{\mathcal{S}}} d(s, \hat{s})<\infty .\notag
	\end{align}
	
	A pair $(R,D)$ is said to be \emph{achievable} if there exists a sequence of $\left(2^{n R}, n\right)$ codes such that	
	\begin{subequations}
		\begin{align}
			&\lim _{n \rightarrow \infty} P_{e}^{(n)}=0, \\ 
			&\varlimsup_{n \rightarrow \infty} D^{(n)} \leq D.
		\end{align} 
	\end{subequations}
	The \emph{capacity-distortion function} is defined as
	\begin{align}
		C(D)=\max\{R:\text{$R$ is achievable for the given $D$} \}.
	\end{align}
	\section{Main results }\label{secp}
	 In this section, we present some properties of the capacity-distortion function $C(D)$. We then derive a multi-letter characterization of $C(D)$ and provide multiple single-letter lower and upper bounds on $C(D)$.
	\begin{lemma}\label{l1}
		The capacity-distortion function $C(D)$  is non-decreasing, concave and continuous for $D\geq D_{\min}=\min\mathbb{E}\big[d(S, \hat{S})\big]$.
	\end{lemma}	 	
	\begin{proof}
		By the definition of $C(D)$, we have  $C(D_1)\ge C(D_2)$ for $D_1>D_2$. To prove the concavity of $C(D)$, we use the time sharing technique. For each block length $n$, assume without loss of generality that $\alpha n$ is an integer for $\alpha \in  [0,1]$ and 
		$R_1<C(D_1), R_2<C(D_2)$. Assume $\left(2^{\alpha n \mathrm{R_1}}, \alpha n\right)$ and $\left(2^{(1-\alpha) n \mathrm{R_2}}, (1-\alpha) n\right)$ are a sequence of codes with rates $R_1$ and $R_2$, respectively. To transmit a message, for the first $\alpha n$ transmissions, the sender transmits a codeword from the first code and for the rest of the transmissions, it transmits a codeword from the second code. Using similar approaches as in~\cite[Theorem 15.3.2]{thomas2006elements}, we can show that the rate-distortion pair $(\alpha {R_1}+(1-\alpha){R_2},\alpha D_1+(1-\alpha)D_2)$ is achievable. According to the definition of capacity-distortion function $C(D)$, we obtain $C(\alpha D_1+(1-\alpha) D_2 )\geq \alpha C(D_1)+(1-\alpha)C(D_2)$.
	\end{proof}
	
Using Lemma~\ref{l1}, we give a multi-letter characterization of the capacity-distortion function $C(D)$ in the following subsection. 
	\subsection{The multi-letter representation of C(D)}\label{mul}
	
	\begin{theorem}\label{th11}
		The capacity-distortion function $C(D)$ satisfies
		\begin{align}\label{m}
			C(D)= \lim_{k \rightarrow \infty}\frac{1}{k}\sup_{P_{X^k}}\big\{  I(X^k ; Y^k | S^k)\big|\mathbb{E}\big[d\big(S^k, \hat{S}^{*k}(Z^k)\big)\big]\leq k D\big\},
		\end{align}
		where $ \hat{S}^{*k}(Z^k)=\big(\hat{S}_1^{*}(Z^k),\hat{S}_2^{*}(Z^k),\cdots,\hat{S}_k^{*}(Z^k)\big)$ and $ \hat{S}_i^{*}(Z^k)=\arg \min_{s^{\prime}}\sum_{s_i}p(s_i|z^k)d(s_i, s^{\prime})$, $ i=1,2,\cdots,k$ is the optimal estimator.   
	\end{theorem}
	\begin{proof}
		We first prove the existence of the limit, then prove achievability and the converse, respectively.
		
		Define that $C_{k}(D)=\frac{1}{k}\sup_{P_X^k}\{  I(X^k ; Y^k | S^k)\big|\mathbb{E}\big[d\big(S^k, \hat{S}^{*k}(Z^k)\big)\big]\leq kD\}$.
		Fix a pmf $p(x^k)$ and a pmf $p(x^l)$, which achieve $C_{k}(D)$ and $C_{l}(D)$ respectively, and the corresponding distortions satisfy $\mathbb{E}\big[d(S^k, \hat{S}^{*k}(Z^k))\big]\leq kD$ and $\mathbb{E}\big[d\big(S^l, \hat{S}^{*l}(Z^l)\big)\big]\leq lD$. Then, fixing the product pmf $p(x^k)p(x^l)$,  we have the corresponding distortion satisfies
		\begin{align}\label{dd1}
			\mathbb{E}\big[d\big(S^{k+l}, \hat{S}^{*k+l}(Z^{k+l})\big)\big]&=\mathbb{E}\big[d\big(S^{k}, \hat{S}^{*k}(Z^{k+l})\big)\big]+\mathbb{E}\big[d\big(S_{k+1}^{k+l}, \hat{S}_{k+1}^{*k+l}(Z^{k+l})\big)\big]\notag \\
			&\stackrel{(a)}{=}\mathbb{E}\big[d\big(S^{k}, \hat{S}^{*k}(Z^{k})\big)\big]+\mathbb{E}\big[d\big(S_{k+1}^{k+l}, \hat{S}_{k+1}^{*k+l}(Z_{k+1}^{k+l})\big)\big]\notag \\
			&\leq (k+l)D,
		\end{align}
		and the corresponding mutual information expression satisfies 
		\begin{align}\label{ikl}
			I(X^{k+l};Y^{k+l}|S^{k+l})
			&=H(Y^{k+l}|S^{k+l})-H(Y^{k+l}|X^{k+l},S^{k+l})\notag \\
			&\stackrel{(b)}{=}H(Y^{k}|S^{k})+H(Y_{k+1}^{k+l}|S_{k+1}^{k+l})-H(Y^{k}|X^{k},S^{k})-H(Y_{k+1}^{k+l}|X_{k+1}^{k+l},S_{k+1}^{k+l})\notag \\
			&= I(X^{k};Y^{k}|S^{k})+I(X_{k+1}^{k+l};Y_{k+1}^{k+l}|S^{k+l})\notag \\
			&=kC_{k}(D)+lC_{l}(D),
		\end{align}
		where $(a)$  follows from the independence of $Z_{k+1}^{k+l}$ and $S^k$ given $Z^{k}$ and $(b)$ follows from the independence of $Y_{k+1}^{k+l}$ and $S^k$ given $Y^{k}$. Thus, according to the definition of $C_{k}(D)$, we get that $(k+l)C_{k+l}(D)\geq kC_{k}(D)+lC_{l}(D)$ for any $k,l\in \mathbb{N}_+$, which means that $kC_{k}(D)$ is superadditive sequence. Further, using the definition of upper and lower limits we get the limit of $C_{k}(D)$ exists and $\lim_{k \rightarrow \infty} C_{k}(D)= \sup C_{k}(D)$.
		
		{\bf Proof of achievability:} 
		Let $p_{X^k}(\cdot)$ be the pmf that achieves $C_{k}(D)$ and also satisfies that $\mathbb{E}\big[d\big(S^k, \hat{S}^{*k}(Z^k)\big)\big]\leq kD.$ For code length $n$, set $n=kt+r,r<k$. Fixing a product pmf $\prod_{i=1}^{t} p(x_{(i-1)k+1}^{ik})p(x^r)$, where $p(x_{(i-1)k+1}^{ik})=p(x^k), i=1,2,\dots ,t$ and $p(x^r)$ satisfies $\mathbb{E}\big[d(S^r, \hat{S}^{*r}(Z^r))\big]\leq rD$. Then, using joint typical decoding at the decoder, we get the average probability of error $P_{e}^{(n)}$ tends to zero if $R<\frac{1}{k}I\left(X^k ; Y^k|S^k\right)-\delta(\epsilon)$ hold, where $\delta(\epsilon)  \rightarrow 0$ when $n \rightarrow \infty$. Similar to the derivation process of (\ref{dd1}), we get the distortion satisfies 
		\begin{align}
			&\frac{1}{n}\mathbb{E}\big[d\big(S^n, \hat{S}^{*n}(Z^n)\big)\big]\notag\\
			=&\frac{1}{n}\Big(\sum_{i=1}^{t} \mathbb{E}\big[d\big(S_{(i-1)k+1}^{ik}, \hat{S}_{(i-1)k+1}^{*ik}(Z^n)\big)\big]+\mathbb{E}\big[d\big(S_{tk+1}^{n}, \hat{S}_{tk+1}^{*n}(Z^n)\big)\big]\Big)\notag\\
			=&\frac{1}{n}\Big(\sum_{i=1}^{t} \mathbb{E}\big[d\big(S_{(i-1)k+1}^{ik}, \hat{S}_{(i-1)k+1}^{*ik}(Z_{(i-1)k+1}^{ik})\big)\big]+\mathbb{E}\big[d\big(S_{tk+1}^{n}, \hat{S}_{tk+1}^{*n}(Z_{tk+1}^{n})\big)\big]\Big)\notag\\
			\leq & \frac{1}{n}(tkD+rD)=D.
		\end{align} 
		
		{\bf Proof of converse:} For any $p(x^k)$ satisfying $\frac{1}{k}\mathbb{E}\big[d\big(S^k, \hat{S}^{*k}(Z^k)\big)\big]\leq D$, we have 
		\begin{align}
			\quad \ k R
			&\leq I\left(W ; Y^k,S^k\right)+k \epsilon_k= I\left(W; Y^k|S^k\right)+k \epsilon_k\notag\\
			&\leq I\left(X^k; Y^k|S^k\right)+k \epsilon_k\stackrel{(a)}{\leq}  kC_k(D)+k \epsilon_k,
		\end{align}
		where $(a)$ follows by the definition of capacity-distortion function $C_k(D)$. Then, we have $R \leq C_k(D)+ \epsilon_k$ and get the conclusion by taking limits on both sides of the inequality.
		
		In the end, we prove that the estimator $\hat{S}_i^{*}(Z^k)$ in Theorem \ref{th11} is the optimal estimator.
		According to the definition of distortion, we have
		\begin{align}
			&\quad \ \mathbb{E}\big[d\big(S^k, \hat{S}^k(Z^k)\big)\big]= \sum_{i=1}^k\mathbb{E}\big[d\big(S_i, \hat{S}_i(Z^k)\big)\big] \notag\\
			&=\sum_{i=1}^k\sum_{z^k}p(z^k)\sum_{\hat{s}_i}p(\hat{s}_i|z^k)\sum_{s_i}p(s_i|z^k)d(s_i, \hat{s}_i) \notag\\
			&\geq\sum_{i=1}^k\sum_{z^k}p(z^k)\min_{s^{\prime}}\sum_{s_i}p(s_i|z^k)d(s_i, s^{\prime}) \notag\\	    
			&\stackrel{(\mathrm{a})}{=}  \sum_{i=1}^k\sum_{z^k}p(z^k)\sum_{s_i}p(s_i|z^k)d(s_i, \hat{s}_i^*) \notag\\
			&= \sum_{i=1}^k\mathbb{E}\big[d\big(S_i, \hat{S}_i^*(Z^k)\big)\big]= \mathbb{E}\big[d\big(S^k, \hat{S}^{*k}(Z^k)\big)\big],
		\end{align}
		where $(a)$ follows by choosing $\hat{S}_i^*(z^k)$ in Theorem \ref{th11}. 
	\end{proof}
	From the above derivation, we conclude that in the bistatic ISAC model under consideration, the reason that prevents the single-letter representation of the capacity-distortion function from being obtained is that the optimal estimator is a sequence estimator. In the monostatic ISAC model, the optimal estimator is one-shot, i.e., the estimator estimates the state $s_i$ in time slot $i$ only depends on $x_i$ and $z_i$, since the estimator naturally aware of the sent message $X$ and the Markov chain $(X^{i-1}, X_{i+1}^n, Z^{i-1}, Z_{i+1}^n,\hat{S}_i) - (X_i,Z_i) - S_i$ holds, which leads to a single-letter representation of the capacity-distortion function.
	
	The supremum in Theorem \ref{th11} is defined over the entire space of joint distributions $P_X^k$. The following theorem shows that we can restrict the optimization variables to all stationary and ergodic stochastic processes.
	\begin{theorem}\label{th2}
		The capacity-distortion function $C(D)$  satisfies
		\begin{align*}
			C(D) = C_{S,E}(D) = \sup_{X'}\big\{  I(X' ; Y' | S')\big|\lim_{k \rightarrow \infty}\frac{1}{k}\mathbb{E}\big[d\big(S^k, \hat{S}^{*k}(Z^k)\big)\big]\leq D\big\},
		\end{align*}
		where the supremum is takev over all staionary and ergodic processes $X'$, $Y' $ is the output process obtained by passing  $X'$ through the bistatic channel model, and $I(X' ; Y' | S')= \lim_{k \rightarrow \infty} \frac{1}{k}  I(X^k ; Y^k | S^k)$.     	
	\end{theorem}
	\begin{proof}
	See Appendix \ref{prth}.
	\end{proof}  

 \subsection{Lower bounds of C(D) }\label{inn}
	 Although the multi-letter characterization of capacity-distortion function is well-defined in the above subsection, it is still not clear how to compute it. Therefore, in this subsection, we propose some single-letter lower bounds on the capacity-distortion function.
		
	In the bistatic ISAC system shown in Fig. \ref{MO1}, the estimator does not have access to the sent messages, resulting in larger estimation errors compared to the monostatic ISAC system, where the estimator knows the sent messages. To mitigate this situation, we leverage communication to assist sensing, meaning that the decoder at the SenRx decodes the sent messages to some extent, thereby facilitating the estimation based on the decoded information. Based on the degree to which communication assists sensing at the SenRx, we propose three \emph{decoding-and-estimation} (DnE) strategies: blind estimation, partial-decoding-based estimation, and full-decoding-based estimation, which lead to three lower bounds on capacity-distortion function.
	\begin{corollary} \label{co33}
		(Blind estimation) The capacity-distortion function $C(D)$ satisfies
		\begin{align}\label{cco3}
			C(D)\geq \max_{P_X}\big\{ I(X ; Y|S)\big|\mathbb{E}\left[d\big(S, \hat{s}^*(Z)\big)\right] \leq D\big\},
		\end{align}
		where the joint distribution of $S X Y Z $ is given by $ P_XP_SP_{YZ|X S}$ and the estimator $\hat{s}^*(z) =\arg \min _{s^{\prime} \in \mathcal{S}} \sum_{s \in \mathcal{S}} P_{S|Z}(s | z) d\left(s, s^{\prime}\right)$.
	\end{corollary} 
	\begin{proof}
		From the proof of Theorem \ref{th11} we obtain $C(D)=\lim_{k \rightarrow \infty} C_{k}(D)= \sup_k C_{k}(D)\geq C_{1}(D)=\max_{P_X}\big\{ I(X ; Y|S)\big|\mathbb{E}\left[d\big(S, \hat{s}^*(Z)\big)\right] \leq D\big\}$, which completes the proof.
	\end{proof}
	In Corollary \ref{co33}, blind estimation is reflected in that the estimator $\hat{s}^*(Z)$ does not decode the sent message and only relies on the received data $Z$. In the following, we apply the partial-decoding-based estimation strategy to obtain a new lower bound.
	\begin{theorem}\label{th1}
		(Partial-decoding-based estimation) The capacity-distortion function $C(D)$  satisfies
		\begin{align}\label{C}
			C(D)\geq \max_{P_{UX}}\big\{  \min\{ I(U;Z)+I(X ; Y | U,S), I(X ; Y | S)\}\big|\mathbb{E}\left[d\big(S, \hat{s}^*(U,Z)\big)\right]\leq D\big\},
		\end{align}
		where $\hat{s}^*(u, z)=\arg \min _{s^{\prime} \in \mathcal{S}} \sum_{s \in \mathcal{S}} P_{S|U Z}(s | u, z) d\left(s, s^{\prime}\right)$. The joint distribution of $S U X Y Z $ is given by $ P_{UX}P_SP_{YZ|X S}$ for some pmf $P_{UX}$, and the cardinality of the auxiliary random variable $U$ satisfies $\left|\mathcal{U}\right| \leq|\mathcal{X}|+1$.
	\end{theorem}
	\begin{proof}
		The proof is mainly divided into two parts: the rate expressions and the state distortion constraint.

		a) Codebook generation. Fix a pmf $p(u) p(x|u)$ such that the expected distortion is less than $ \big( D /(1+\epsilon) \big)$ for a small positive number $\epsilon>0$, where $D$ is the desired distortion. Randomly and independently generate $2^{n R_0}$ sequences $u^n\left(w_0\right), w_0 \in\left[1: 2^{n R_0}\right]$, each according to $\prod_{i=1}^n p_{U}(u_i)$. For each $w_0 \in$ $\left[1: 2^{n R_0}\right]$, randomly and conditionally independently generate $2^{n R_1}$ sequences $x^n\left(w_0, w_1\right)$, $w_1 \in\left[1: 2^{n R_1}\right]$, each according to $\prod_{i=1}^n p_{X|U}\big(x_i| u_i(w_0)\big)$.
		
		b) Encoding. To send $\left(w_0, w_1\right)$, transmit $x^n\left(w_0, w_1\right)$.
		
		c) Decoding. Decoder at the ComRx declares that $(\hat{w}_{01},\hat{w}_1)$ is sent if it is the unique message pair such that $\big({u^n(\hat{w}_{01})}, x^n(\hat{w}_{01}, \hat{w}_1), y^n, s^n\big) \in \mathcal{T}_\epsilon^{(n)}(P_{UXYS})$, where $\mathcal{T}_\epsilon^{(n)}$ refers to jointly $\epsilon$-typical sequences\footnote{The definition of the joint typical sequence here refers to the definition in \cite{el2011network}, i.e., robust typicality, which is convenient to use the corresponding theorem, for example, the conditional typicality lemma \cite[Lemma 2.5]{el2011network}.}; otherwise it declares an error. The SenRx declares that $\hat{w}_{02}$ is sent if it is the unique message such that $\left(u^n\left(\hat{w}_{02}\right)\right.$, $\left.z^n\right) \in \mathcal{T}_\epsilon^{(n)}(P_{UZ})$; otherwise it declares an error. 
		
		d) Estimation. Assuming that the sensing channel output is $Z^n=z^n$ and the decoded information is $\hat{U}^n=\hat{u}^n$, then the one-shot estimator gives the estimation of the state sequences
		\begin{align*}  
			\hat{S}^n=\big(\hat{s}^*(\hat{u}_1,z_1), \hat{s}^*(\hat{u}_2,z_2), \ldots, \hat{s}^*(\hat{u}_n,z_n)\big) .
		\end{align*}
		
		e) Analysis of the probability of error.
		
		Assume, without loss of generality, that $\left(W_{0}, W_1\right)=(1,1)$ is sent. The average probability of error is $ P_{1e}^{(n)}=\mathrm{P}\big\{(\hat{W}_{01}, \hat{W}_1) \neq\left(1, 1\right) \text { or }\hat{W}_{02} \neq1\big\}$.
		Applying the jointly typical decoding at the decoder and following similar arguments as superposition coding inner bound, we obtain that $P_{1e}^{(n)}$ tends to zero if $R_0<I\left(U ; Z\right)-\delta(\epsilon)$, $R_1<I\left(X ; Y|U,S\right)-\delta(\epsilon)$, and $R_0+R_1<I\left(X ; Y|S\right)-\delta(\epsilon)$ hold, where $\delta(\epsilon)  \rightarrow 0$ when $n \rightarrow \infty$. 
		
		f) Analysis of the expected distortion 
		
		Define the correct decoding event as $\mathcal{A}=\big\{(\hat{W}_{01}, \hat{W}_1)=(1, 1) \text { and }\hat{W}_{02}=1\big\}$ and the complement of $\mathcal{A}$ as $\mathcal{A}^c$. The expected distortion (averaged over the random codebook, state and channel noise) can be upper bounded by
		\begin{align} \label{d11}
			D^{(n)} &=\frac{1}{n} \sum_{i=1}^n \mathrm{E}\big[d(S_i, \hat{S}_i)\big] \notag\\
			&=\frac{1}{n} \operatorname{P}(\mathcal{A}^c)\sum_{i=1}^n \mathrm{E}\big[d(S_i, \hat{S}_i) \big| \mathcal{A}^c\big]+ 
			\frac{1}{n} \operatorname{P}(\mathcal{A})\sum_{i=1}^n \mathrm{E}\big[d(S_i, \hat{S}_i) \big| \mathcal{A}\big]   \notag\\
			&\leq d_{\max } P_{1e}^{(n)} +\frac{1}{n} \sum_{i=1}^n \mathrm{E}\big[d(S_i, \hat{S}_i) \big| \mathcal{A}\big](1-P_{1e}^{(n)}).
		\end{align}
		
		According to the decoding principle, we have $ \big(U^n(1),X^n(1,1),
		Y^n,S^n\big) \in \mathcal{T}_\epsilon^{(n)}\left( P_{UX}P_S P_{Y | S X}\right)$ and $	\big(U^n(1), Z^n\big) \in \mathcal{T}_\epsilon^{(n)}\left(P_{U} P_{Z|U}\right)$ when decoded correctly.
		Furthermore, according to the fact that $\hat{S}_i=\hat{s}^*\left(U_i,Z_i\right)$ and the conditional typicality lemma \cite[Lemma 2.5]{el2011network}, we get for every $\epsilon^{\prime}>\epsilon$, $P\Big(\big(S^n, U^n(1), \hat{S}^n\big) \in \mathcal{T}_{\epsilon^{\prime}}^{(n)}(P_{S U \hat{S}})\Big)=1-\eta$, 
		where $P_{S U \hat{S}}$ denotes the joint marginal distribution of $P_{S U X Z \hat{S}}(s, u,x, z, \hat{s})= P_{UX}(u,x)\cdot P_S(s)\cdot $ $ P_{Z | S X}(z |s, x)\cdot \chi\big( \hat{s}=\hat{s}^*(u,z)\big)$, $\chi(\cdot) $ is the indicator function, $\eta\in(0,1)$ and $\lim_{n \to \infty}\eta =0  $. Then, we define  $(S^n, U^n(1), \hat{S}^n) \in \mathcal{T}_{\epsilon^{\prime}}^{(n)}(P_{S U \hat{S}})$ as event $\mathcal{B}$ and the complement of $\mathcal{B}$ as $\mathcal{B}^c $. From\cite[Typical Average Lemma]{el2011network} it follows that
		\begin{align} \notag\label{d22}
			&\quad \ \varlimsup_{n \rightarrow \infty} \frac{1}{n} \sum_{i=1}^n \mathbb{E}\big[d(S_i, \hat{S}_i)\big| \mathcal{A}\big]\notag\\
			&= \varlimsup_{n \rightarrow \infty} \frac{1}{n} (1-\eta)\sum_{i=1}^n \mathbb{E}\big[d(S_i, \hat{S}_i)\big|\mathcal{A}, \mathcal{B}\big] +\varlimsup_{n \rightarrow \infty} \frac{1}{n} \eta \sum_{i=1}^n \mathbb{E}\big[d(S_i, \hat{S}_i)\big|\mathcal{A}, \mathcal{B}^c\big]\notag\\
			&\leq \varlimsup_{n \rightarrow \infty} (1-\eta)(1+{\epsilon^{\prime}}) \mathbb{E}[d(S, \hat{S})] + \eta d_{\max}= (1+{\epsilon^{\prime}}) \mathbb{E}[d(S, \hat{S})]
		\end{align}
		for $(S, \hat{S})$ following the joint marginal distribution of $P_{S U Z \hat{S}}$ defined above. Thus, we obtain from (\ref{d11}) and (\ref{d22}) that $\varlimsup_{n \rightarrow \infty} D^{(n)}\leq (1+{\epsilon^{\prime}}) \mathbb{E}[d(S, \hat{S})]$.
		Taking ${\epsilon^{\prime}} \rightarrow \epsilon$, we conclude that the error probability vanishes and the distortion constraint holds when~\eqref{C} holds. Besides, according to the cardinality bounding technique in \cite[Appendix C]{el2011network} and considering the state variable $S$, we get the cardinality of the auxiliary random variable $U$ satisfies $\left|\mathcal{U}\right| \leq  |\mathcal{X}|+1$.
	\end{proof}
	Note that for  $S U X Y Z $ distributed as $ P_{UX}P_SP_{YZ|X S}$,  $\hat{s}^*(u, z)$
	is the optimal one-shot state estimator, which means that the estimator estimates the state $s_i$ in time slot $i$ only as a function of $u_i$ and $z_i$. In addition, $\hat{s}^*(u, z)$ can be interpreted as a minimizer of the penalty function on the distortion measure $d$ and  $P_{S|U Z}(s|u, z)$ is the posterior probability of $S$ when $(U,Z)$ is known. In particular, when the distortion measure $d$ is the Hamming distance, $\hat{s}^*(u, z)$ is a maximum a posteriori probability estimate. The proof on the optimality is given in Appendix~\ref{prl1}. 
 
	\begin{remark}
		Referring to the broadcast channel model, we can obtain a more general lower bound using the idea of Marton's inner bound \cite[Theorem 8.4]{el2011network}. However, since its expression form is more complicated and the corresponding proof follows similar steps to the proof above and thus the details of which are omitted.
	\end{remark}

	Applying the full-decoding-based estimation strategy, where the SenRx fully decodes the sent message to assist estimation, we obtain the following lower bound on the capacity-distortion function.
	\begin{corollary} \label{co2}
		(Full-decoding-based estimation) The capacity-distortion function $C(D)$ satisfies
		\begin{align}\label{cco1}
			C(D)\geq \max_{P_X}\big\{ \min \{I(X ; Y|S),I(X ; Z)\}\big|\mathbb{E}\left[d\big(S, \hat{s}^*(X,Z)\big)\right] \leq D\big\},
		\end{align}
		where the joint distribution of $S X Y Z $ is given by $ P_XP_SP_{YZ|X S}$ and the optimal estimator $\hat{s}^*(x, z) =\arg \min _{s^{\prime} \in \hat{\mathcal{S}}} \sum_{s \in \mathcal{S}} P_{S|X Z}(s | x, z) d\left(s, s^{\prime}\right)$.
	\end{corollary} 
	\begin{proof}
		Similar to the proof of Theorem \ref{th1}, the conclusion is obtained by letting the SenRx decode the sent information $X$ and applying the optimal estimator $\hat{S}^n=\big(\hat{s}^*(\hat{x}_1,z_1), \hat{s}^*(\hat{x}_2,z_2), \ldots, \hat{s}^*(\hat{x}_n,\\z_n)\big)$.
	\end{proof} 
 
	\begin{remark}			
			From the expression in Theorem \ref{th1}, we observe that when $U=\emptyset$, the result of Theorem \ref{th1} reduces to that of Corollary \ref{co33}, and when $U=X$, the result of Theorem \ref{th1} reduces to that of Corollary \ref{co2}, where $U$ is viewed as the amount of information decoded to assist estimation.  This observation indicates that the partial-decoding-based estimation strategy contains both blind estimation and full-decoding-based estimation strategies as the special cases.
	\end{remark}
		\begin{remark}\label{maxdis}
		Note that we adopt the average expected distortion defined in \eqref{D} as the distortion metric, which may not be more meaningful than the maximum distortion in some scenarios. However, they are equivalent in the case under consideration, i.e., all the single-letter bounds presented in the paper also apply to the maximum distortion measure. For converse statements, the maximum is more restrictive than the average. If the converse holds for the average, i.e., no code exists with a certain average performance in terms of error probability and distortion, then also no code exists with the maximum performance, which means the converse holds for the maximum. Therefore, in the following, we primarily analyze the lower bound (achievability) based on the average metric. Specifically, based on the model in Fig. \ref{MO1}, we define that a maximum expected distortion as 
		\begin{align*}
			\Delta_n^{\max} \triangleq \max_{w} \Delta_n\big(x^n(w)\big)=\max _{w }\frac{1}{n}\mathbb{E}\big[d\big(S^n,\hat{S}^n(Z^n)\big)|W=w\big].
		\end{align*}		
		To reconcile the maximum distortion and average distortion, and to bring the input distribution $P_X$ into the picture, we consider the type of a sequence. For a sequence $x^n \in \mathcal{X}^n$, its type is a distribution on $\mathcal{X}$ defined as $\pi(a|x^n) \triangleq \frac{1}{n}\sum_{i=1}^{n} \mathbbm{1}[x_i=a]$.
		Going back to the expected distortion given a sequence $X^n =x^n$, we have
		\begin{align}\label{maxd}
			&\Delta_n(x^n)=  \frac{1}{n}\mathbb{E}\big[d\big(S^n,\hat{S}^n(Z^n)\big)|W=w\big]\stackrel{(a)}{=}\frac{1}{n}\sum_{i=1}^n\mathbb{E} \big[  d \big(S_i, \hat{S}_i({Z_i}) \big)  \big| X_i = x_i \big]\notag\\
			= &\frac{1}{n}\sum_{i=1}^n\sum_{s_i \in \mathcal{S}} \sum_{z_i \in \mathcal{Z}} P_{Z, S|X}(z_i,s_i | x_i) d\big(s_i, \hat{s}_i(z_i) \big)
			= \frac{1}{n}\sum_{i=1}^n\sum_{s,z} P_{Z, S|X}(z,s | x_i) d\big(s, \hat{s}(z) \big)\notag\\
			= &\frac{1}{n}\sum_{i=1}^n\mathbb{E}\big[d\big(S, \hat{S}(Z)\big) |X_i= x_i\big]
			=\frac{1}{n} \sum_{i=1}^n \sum_{a \in \mathcal{X}} \mathbbm{1}[x_i = a] \mathbb{E} \big[ d \big( S, \hat{{S}}(Z) \big) \big| X_i = a \big]\notag\\
			= &\sum_{a \in \mathcal{X}} \pi(a|x^n) \mathbb{E} \big[ d \big( S, \hat{{S}}(Z) \big) \big| X = a \big]
			\stackrel{(b)}{\to}  \sum_{a \in \mathcal{X}} P_X(a) \mathbb{E} \big[ d \big( S, \hat{{S}}(Z) \big) \big| X = a \big]
			= \mathbb{E} \big[ d \big( S, \hat{{S}}(Z) \big) \big],
		\end{align}
		where $(a)$ holds by the fact that all estimators are one-shot in our lower bounds and $(b)$ holds when $x^n$ belongs to the robust typical set, which is defined as $\mathcal{T}_{\epsilon}^{(n)}(P_X) \triangleq \big\{ x^n \in \mathcal{X}^n : |\pi(a|x^n) - P_X(a)| \leq \epsilon P_X(a) \text{ for all } a \in \mathcal{X} \big\}$.
		Moreover, if all codewords in the codebook have a type which is close enough to $P_X$, then the above will hold for every codeword (i.e., every message).
		
		Comparing \eqref{D} and \eqref{maxd}, we have that the maximum expected distortion and the average expected distortion are essentially the same. Therefore, for the model we consider, the average expected distortion is valid and sufficient.
	\end{remark}

	\subsection{Upper bounds of C(D) }\label{out}
	In this subsection, we propose three upper bounds of the capacity-distortion function with the characterization of a single letter.
	
	An upper bound can be obtained by introducing a genie to the system that feeds the sent message to the estimator, as shown in Theorem \ref{tho1} below.
	
	\begin{theorem} \label{tho1}
		(Upper bound 1) The capacity-distortion function $C(D)$ satisfies
		\begin{align*}
			C(D)\leq C_o(D)=\max_{P_X}\big\{  I(X ; Y |  S)\big|\mathbb{E}\big[d\big(S, \hat{S}^*(X,Z)\big)\big]\leq D\big\},
		\end{align*}
		where the joint distribution of $S X Y Z $ is given by $ P_XP_SP_{YZ|X S}$ and the optimal estimator $\hat{s}^*(x, z) =\arg \min _{s^{\prime} \in \hat{\mathcal{S}}} \sum_{s \in \mathcal{S}} P_{S|X Z}(s | x, z) d\left(s, s^{\prime}\right)$.
	\end{theorem}

	
	\begin{proof}
		When the information sent is revealed (via a genie) to the estimator, the model is similar to that of the monostatic ISAC systems  in \cite {ahmadipour2022information}, thus the details of proof is omitted.
	\end{proof}
	
	\begin{remark}		
		Note that when the SenRx is statistically stronger than the ComRx, i.e., the channel to the ComRx is physically degraded or stochastically degraded with respect to the channel to the SenRx, we get that if the ComRx can decode the message sent, i.e., $R\leq I(X ; Y |  S)$, then the SenRx also can decode the message sent because of $R\leq I(X ; Y |  S)\leq I(X;Z)$. In this case, the estimator obtains the sent message and thus the corresponding capacity-distortion function $C(D)=C_o(D)$.  
	\end{remark}
	
	In the following, we derive a new upper bound by introducing a genie-aided estimator, where the genie supplies the estimator with some prior information instead of the sent message.
	
		\begin{theorem} \label{tho2}
		(Upper bound 2) The capacity-distortion function $C(D)$ satisfies
		\begin{align}\label{cdup2}
			C(D)\leq \max_{P_{UX}}\min&\big\{  I(X;Y|S), I(X ; Y | U, S)+I(U;Z),I(X ; Y,Z | U,V, S)\notag\\
			&+I(U,V;Z)\big|\mathbb{E}\big[d\big(S, \hat{S}^*(U,V,Z)\big)\big]\leq D\big\},
		\end{align}
		where the joint distribution of $U V S X Y Z $ is given by $ P_{UV}P_{X|UV}P_SP_{YZ|X S}$, the estimator $\hat{s}^*(u,v,\\ z)=\arg \min _{s^{\prime} \in \mathcal{S}} \sum_{s \in \mathcal{S}} P_{S|U V Z}(s | u,v, z) d\left(s, s^{\prime}\right)$, and the cardinalities of the auxiliary random variables $U$ and $V$ satisfy $\left|\mathcal{U}\right| \leq|\mathcal{X}|+2$ and $\left|\mathcal{V}\right| \leq|\mathcal{U}||\mathcal{X}|+1$, respectively.
	\end{theorem}
	\begin{proof}
		Based on  Fano's inequality and Csiszar sum identity \cite{el2011network},  we get the rate expression in \eqref{cdup2} by taking the auxiliary random variables as $U_i=\left(Z_{i+1}^n,Y^{i-1},S^{i-1}\right)$ and $V_i=\left(Z^{i-1}\right)$. See Appendix \ref{prth5} for details.
	
		In the following, we consider the distortion constraint. For each estimator satisfying $D\geq \frac{1}{n} \sum_{i=1}^n \mathbb{E}\big[d\big(S_{i}, \hat{S}_i( Z^{n})\big)\big]$, we define a genie-aided estimator ${\hat{S}}'_i(U_i,V_i, Z^{n})$, whose corresponding distortion is less than that of estimator ${\hat{S}}_i(Z^{n})$. Such an estimator exists since more information is exploited. Then, for each genie-aided estimator ${\hat{S}}'_i(U_i,V_i, Z^{n})$, we construct a new genie-aided estimator ${\hat{S}}''_i(U_i,V_i, Z_i)$, which  satisfies ${\hat{S}}'_i(U_i,V_i, Z^{n})={\hat{S}}''_i(U_i,V_i, Z_i)$ due to the composition of the auxiliary random variable identification  $U_i,V_i$. Thus, we have
			\begin{align*}			
				D&\geq \frac{1}{n} \sum_{i=1}^n \mathbb{E}\big[d\big(S_{i}, \hat{S}_i( Z^{n})\big)\big]\notag\\
				&\geq \frac{1}{n} \sum_{i=1}^n \mathbb{E}\big[d\big(S_{i}, {\hat{S}}'_i( U_i,V_i,Z^{n})\big)\big] \notag\\
				&=\frac{1}{n} \sum_{i=1}^n \mathbb{E}\big[d\big(S_{i}, {\hat{S}}''_i( U_i,V_i,Z_i)\big)\big]\notag\\
				&\geq\frac{1}{n} \sum_{i=1}^n \mathbb{E}\big[d\big(S_{i}, \hat{s}^*( U_i,V_i,Z_i)\big)\big].
			\end{align*}
			Recall that the definition of the estimator in Theorem \ref{th1}, $\hat{s}^*\big((U_i,V_i,i), Z_i\big)=\hat{s}^*\left(U_i,V_i, Z_i\right)$ for $i=1,\cdots,n$.  We introduce a time-sharing random variable $Q \sim$ Unif $[1: n]$ independent of $\left(W, X^n, Y^n, Z^n, S^n\right)$ and define that $U=\left(Q, U_Q\right), V=\left(Q, V_Q\right), X=X_Q, Y=Y_Q$, $Z=Z_Q$ and $S=S_Q$. Using the law of total expectation and the definitions of $Q, S, U,V,Z$ above, we have 
			\begin{align}	\label{dcon1n}
				&\frac{1}{n} \sum_{i=1}^n \mathbb{E}\big[d\big(S_{i}, \hat{s}^*( U_i,V_i,Z_i)\big)\big]	=\frac{1}{n} \sum_{i=1}^n \mathbb{E}\Big[d\Big(S_i, \hat{s}^*\big((U_i,V_i,i), Z_i \big)\Big)\Big|Q=i\Big]\notag\\		=&\mathbb{E}\bigg[\mathbb{E}\Big[d\Big(S_{Q}, \hat{s}^*\big((U_Q,V_Q,Q), Z_Q \big)\Big)\Big]\Big|Q\bigg]=\mathbb{E}\big[d\big(S, \hat{s}^*(U,V, Z)\big)\big],
			\end{align}
			which completes the proof.			
	\end{proof}
Note that when the channel to SenRx is degraded with respect to the channel to ComRx, we have the following result.
		\begin{theorem}\label{cob1}
			The capacity-distortion function $C(D)$ for the case where $(X,S) - Y - Z$ forms a Markov chain satisfies 
			\begin{align}\label{sendeg}
				C(D)= \max_{P_{UX}}\big\{ I(X ; Y | U, S)+I(U;Z)\big|\mathbb{E}\big[d\big(S, \hat{S}^*(U,Z)\big)\big]\leq D\big\},
			\end{align}
			where $\hat{s}^*(u, z)$ is the same as defined in Theorem \ref{th1}. The joint distribution of $S U X Y Z $ is given by $ P_{UX}P_SP_{YZ|X S}$ for some pmf $P_{UX}$, and the cardinality of the auxiliary random variable $U$ satisfies $\left|\mathcal{U}\right| \leq|\mathcal{X}|+1$. 
		\end{theorem}
	\begin{proof}
		Since $(X,S) - Y - Z$ forms a Markov chain, we have  $I\left(U ; Y|S\right) \geq I\left(U ; Y\right) \geq I\left(U ; Z\right)$ holds for all $p(u, x)$. Hence, we obtain that 
		 \begin{align*}
		 	I\left(U ; Z\right)+I\left(X ; Y | U ,S\right) \leq I\left(U ; Y|S\right)+I\left(X ; Y | U,S\right)=I\left(X ; Y|S\right),
		 \end{align*}
		  and 
		  \begin{align*}
		  I(X ; Y,Z | U,V, S)+I(U,V;Z)=I(X ; Y| U,V, S)+I(U,V;Z):=I(X ; Y| \tilde{U}, S)+I(\tilde{U};Z).
		\end{align*} 
	Also, we have $\mathbb{E}\big[d\big(S, \hat{S}^*(U,V,Z)\big)\big]=\mathbb{E}\big[d\big(S, \hat{S}^*(\tilde{U},Z)\big)\big]$. Therefore, we get that 
		\begin{align*}
		C(D)\leq \max_{P_{ \tilde{U}X}}\big\{ I(X ; Y| \tilde{U}, S)+I( \tilde{U};Z)\big|\mathbb{E}\big[d\big(S, \hat{S}^*(\tilde{U},Z)\big)\big]\leq D\big\}
		\end{align*}
	according to Theorem \ref{tho2}. Furthermore, we have 
	\begin{align*}
		C(D)\geq \max_{P_{ UX}}\big\{ I(X ; Y| U, S)+I( U;Z)\big|\mathbb{E}\big[d\big(S, \hat{S}^*(U,Z)\big)\big]\leq D\big\}
	\end{align*}
	according to Theorem \ref{th1}, which indicates the upper bound 2 coincides with the lower bound in Theorem \ref{th1}. Thus, we obtain the capacity-distortion function $C(D)$, as shown in \eqref{sendeg}.
	\end{proof}
	\begin{remark}
Theorem \ref{cob1} implies that the optimality of the partial-decoding-based estimation strategy achieved by superposition coding in the degraded case. In addition, the result in Theorem \ref{cob1} also holds when the Markov chain $X - (Y, S) - Z$ holds.
	\end{remark}
	 Referring to the lossy source coding for this ISAC system, we get the following upper bound.
	
	\begin{theorem} \label{tho}
		(Upper bound 3) The capacity-distortion function $C(D)$ satisfies
		\begin{align}
			C(D)\leq \max_{{P_{UX}}:I(U,S ; Z) \geq R_S(D) } \big\{I(X ; Y|U,S)+I(U,S ; Z)\big\}-R_S(D),
		\end{align}
		where $ R_S(D)=\min_{P(\hat{S}|S):\mathbb{E}d(S, \hat{S})\leq D}I(S ; \hat{S})$.
	\end{theorem} 
	\begin{proof}
		According to Fano's inequality \cite{el2011network}, we get that
		\begin{align}
			& \quad \ n \left(R + R_{s}(D)\right)-n \epsilon_n\leq I\left(W ; Y^n|S^n\right)+I(S^n ; \hat{S}^n) \stackrel{(a)}\leq I\left(W ; Y^n|S^n\right)+I\left(S^n ; Z^n\right) \notag\\
			& = \sum_{i=1}^nI\left(W; Y_i|Y^{i-1},S^{n}\right) + I\left(S^n; Z_i|Z_{i+1}^n\right) \leq \sum_{i=1}^nI\left(W,Z_{i+1}^n; Y_i|Y^{i-1},S^{n}\right) + 	I\left(S^n,Z_{i+1}^n; Z_i\right)\notag\\
			& = \sum_{i=1}^nI\left(W,Z_{i+1}^n; Y_i|Y^{i-1},S^{n}\right) + 	I\left(S^n,Z_{i+1}^n,Y^{i-1}; Z_i\right)-I\left(Y^{i-1}; Z_i|S^n,Z_{i+1}^n\right)\notag\\
			& \stackrel{(b)}{=} \sum_{i=1}^nI\left(W; Y_i|Y^{i-1},S^{n},Z_{i+1}^n\right) + 	I\left(S^n,Z_{i+1}^n,Y^{i-1}; Z_i\right)\notag\\
			& \leq  \sum_{i=1}^nI\left(X_i; Y_i|Y^{i-1},S^{n},Z_{i+1}^n\right) + 	I\left(S^n,Z_{i+1}^n,Y^{i-1}; Z_i\right)\notag\\
			& =  \sum_{i=1}^nI\left(X_i; Y_i|S_i,U_i\right) + 	I\left(S_i,U_i; Z_i\right)
		\end{align}
		where $(a)$ follows by the Markov chain $S^n-Z^n-\hat{S}^n$, $(b)$ follows by the Csiszar sum identity, and the auxiliary random variable identification $U_i=\left( Y^{i-1},Z_{i+1}^n,S^{i-1},S_{i+1}^n\right)$. 		
		
		Furthermore, similar to the converse proof in lossy source coding, we have $\sum_{i=1}^n I(S_i; \hat{S}_i)\leq I(S^n; \hat{S}^n)$. And since the Markov chain $S^n-Z^n-\hat{S}^n$ and the composition of auxiliary random variable $U_i$, we get 
		\begin{align*}
		 \sum_{i=1}^n I(S_i; \hat{S}_i)\leq I(S^n; \hat{S}^n)\leq I(S^n; Z^n)\leq \sum_{i=1}^n I(S_i,U_i; Z_i),
		\end{align*}
		which completes the proof.
	\end{proof} 
	\subsection{Capacity-distortion region for a bistatic ISAC broadcast channel model}

\begin{figure}[!h]
		\centering
		\includegraphics[width=4.5in]{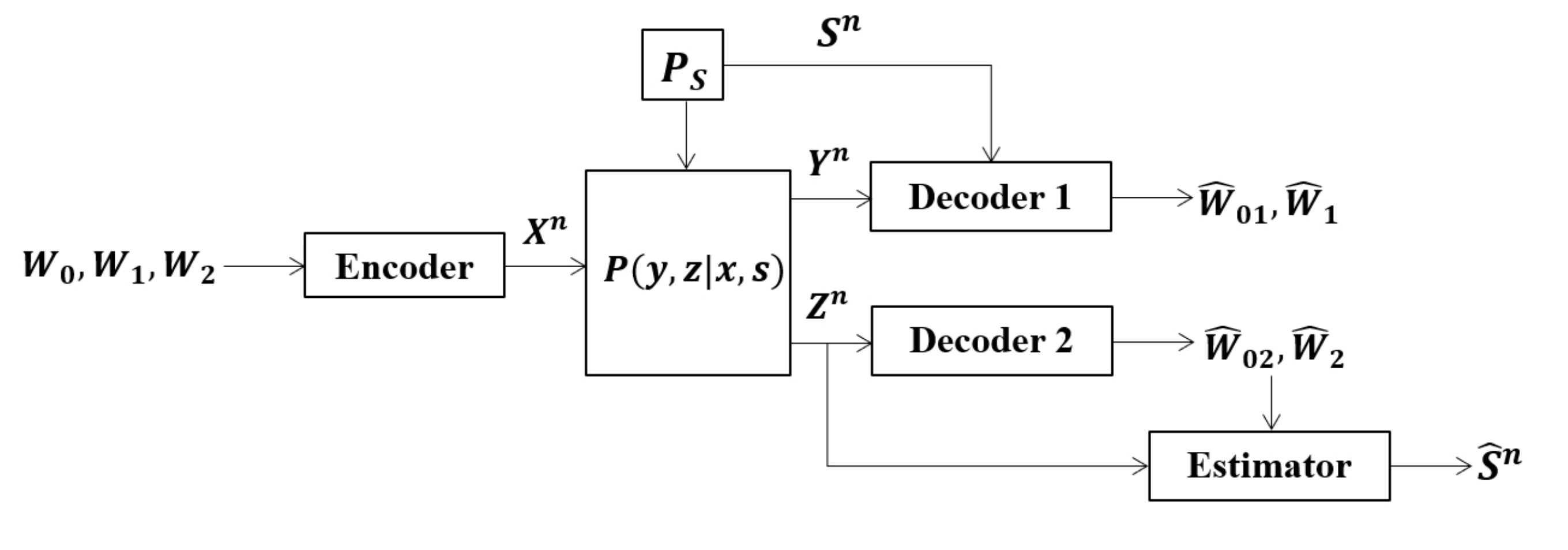}
		\caption{Two-receiver SDMBC model.}
		\label{2}
	\end{figure}

	In this subsection, we consider the bistatic ISAC broadcast systems, where the transmitter sends the message to the ComRx and the SenRx, the ComRx and the SenRx decode the information, and SenRx also needs to estimate the state of the channel. We propose a two-receiver state-dependent memoryless broadcast channel (SDMBC) model $\left(\mathcal{X}, \mathcal{S}, p\left(y,z|x,s\right), \mathcal{Y} \times \mathcal{Z}\right)$, which consists of four finite sets $\mathcal{X}, \mathcal{S}, \mathcal{Y}, \mathcal{Z}$ and a collection of pmf $p\left(y,z|x,s\right)$ on $\mathcal{Y} \times \mathcal{Z}$. As shown in Fig. \ref{2}, the ISAC Tx sends a message $W=(W_0,W_1,W_2)$, where $W_0$ is the common message to the ComRx and the SenRx, $W_1$ and $W_2$  are the private message to the ComRx and the SenRx, respectively.

	A $\left(2^{n R_0}, 2^{n R_1},2^{n R_2}, n\right)$ code for SDMBC consists of
	
	1) three message sets $\left[1: 2^{n R_0}\right] $, $\left[1: 2^{n R_1}\right] $, and $\left[1: 2^{n R_2}\right]$;
	
	2) an encoder that assigns a codeword $x^n\left(w_0, w_1, w_2\right)$ to each message pair $\left(w_0, w_1, w_2\right) \in\left[1: 2^{n R_0}\right] \times\left[1: 2^{n R_1}\right]\times\left[1: 2^{n R_2}\right] $;
	
	3) two decoders, where Decoder 1 assigns an estimate $\left(\hat{w}_{01}, \hat{w}_1\right) \in\left[1: 2^{n R_0}\right] \times\left[1: 2^{n R_1}\right]$ to each $(y^n,s^n)$ and Decoder 2 assigns an estimate $\left(\hat{w}_{02}, \hat{w}_2\right) \in\left[1: 2^{n R_0}\right] \times\left[1: 2^{n R_2}\right]$ to each $z^n$;
	
	4) a state estimator $h_n: \mathcal{Z}^n  \rightarrow \mathcal{S}^n  $, where Decoder 2 estimates $\hat{s}^n=h_n(z^n )$.  
	
	We assume that the message pair $\left(W_0, W_1, W_2\right)$ is uniformly distributed over $\left[1: 2^{n R_0}\right] \times$ $\left[1: 2^{n R_1}\right] \times\left[1: 2^{n R_2}\right]$ and declare an error event as at least one decoding error occurs in Decoder 1 and Decoder 2. Thus, the average probability of error is given by $P_{2e}^{(n)}=\mathrm{Pr}\big\{(\hat{W}_{01}, \hat{W}_1) \neq\left(W_0, W_1\right) \text { or }(\hat{W}_{02}, \hat{W}_2) \neq\left(W_0, W_2\right)\big\}$.
	
	A rate-distortion tuple $(R_0,R_1,R_2,D)$ is said to be {\em achievable} if there exists a sequence of $\left(2^{n R_0}, 2^{n R_1}, 2^{n R_2}, n\right)$ codes such that	$\lim _{n \rightarrow \infty} P_{2e}^{(n)}=0,\varlimsup_{n \rightarrow \infty} D^{(n)} \leq D$, where $D$ is the desired maximum distortion. The \emph{capacity-distortion region} of a SDMBC model is given by the closure of the union of all achievable rate-distortion tuples $(R_0,R_1,R_2,D)$. 
	
	For simplicity of presentation, we focus the discussion on the private-message capacity-distortion region, i.e., the capacity-distortion region when $R_0 = 0$, as that in broacast channel and the results can be extended to the case of both common and private messages. We denote the capacity-distortion region corresponding to such a system as $C_{BC}$. The following theorem gives the capacity-distortion region of this system when the ComRx is statistically stronger than the SenRx.
	\begin{theorem}\label{th3}
		The capacity-distortion region $C_{BC}$ for the case where $(X,S) - Y - Z$ forms a Markov chain  satisfies 
		\begin{equation}\label{degrac}
			C_{BC}= \bigcup_{P_{UX}}\left\{
			\begin{array}{lc}
				(R_1,R_2,D) \in \mathbb{R}_+^3:&  \\
				R_2\leq I(U;Z) &  \\
				R_1+R_2\leq I(X ; Y | U,S)+I(U;Z) & \\
				\mathbb{E}[d\big(S, \hat{s}^*(U,Z)\big)\big] \leq D 
			\end{array}
			\right\}
		\end{equation}
		where $\hat{s}^*(u, z)$ is the same as defined in Theorem \ref{th1}. The joint distribution of $S U X Y Z $ is given by $ P_{UX}P_SP_{YZ|X S}$ for some pmf $P_{UX}$, and the cardinality of the auxiliary random variable $U$ satisfies $\left|\mathcal{U}\right| \leq|\mathcal{X}|+1$. 
	\end{theorem}
	\begin{proof}
Combined the lower bound in Theorem \ref{th1} and the upper bound in Theorem \ref{tho2}, we get the capacity-distortion region $C_{BC}$ as shown in \eqref{degrac} by choosing the auxiliary variable $U_i=\left(W_0,Z_{i+1}^n,Y^{i-1},S^{i-1},Z^{i-1}\right)$. See Appendix \ref{prth7} for details.
	\end{proof}
	Theorem \ref{th3} shows that in the case when $Z$ is degraded with respect to $Y$,  the optimal one-shot estimator outperforms estimators that depend on the output sequence. Intuitively speaking, in the case when $Z$ is degraded with respect to $Y$, $Y$ contains more information about $X$ than $Z$. Compared to an estimator that only relies on the output sequence $Z^n$, the estimator in Theorem \ref{th3} uses not only $Z$ but also part of the information of $Y$ in the auxiliary variable $U$, leading to better estimates. In fact, using similar arguments, we can show that Theorem \ref{th3} also holds in the case where the Markov chain  $ X - (Y,S) - Z$  holds. 
	
	\begin{remark}
	Since both capacity-distortion function and capacity-distortion region in this paper depend on the channel conditional pmf only through the conditional marginal pmfs, our capacity-distortion results for the physically degraded case defined by Markov chain are the same as those for stochastically degraded case.
\end{remark}
 \begin{remark} 
		Note that the rate expression in Theorem \ref{th3} differs from that of the general degraded broadcast channel. The change in the composition of the auxiliary variable makes the converse proof not obtainable as that in the degraded broadcast channel. In addition, the equivalence of the rate region for the more capable broadcast channel no longer holds.
	\end{remark}
	\section{Numerical Examples}\label{secex}
	
	In this section, we provide two examples to illustrate explicitly the results in the previous section.
	
\subsection{Example 1}
		We consider a binary channel model, in which the signals received by the ComRx and the SenRx are given by
		$Z=Y=S\oplus X$  with binary alphabets $\mathcal{X}=\mathcal{S}=\mathcal{Y}=\{0,1\}$ and the state $S$ is Bernoulli$(q)$, for $q \in(0,1)$. The Hamming distortion measure $d(s, \hat{s})=s \oplus \hat{s}$ is employed to characterize the sensing accuracy.     		

 In the following, 
we apply the theoretical results in section \ref{secp} to derive the capacity-distortion function.
Let $\bar{e}=1-e$, $P(X=0)=p\in [0,1]$, and $e\ast p =e\bar{p}+\bar{e}p$. Besides, ${\rm{H}}_2(p)=-p \log_2 p -(1-p) \log_2 (1-p)$ is the binary entropy function and $\rm{H}_2^{-1}(\cdot)$ refers to the inverse function of the binary entropy function $\rm{H}_2(\cdot)$ with the variable taking the value in $[0,1/2]$. It is seen from the channel model in Example 1 that $Z$ is a degraded form of $Y$, i.e., the Markov chain $(X,S) - Y - Z$ holds. It follows from Theorem \ref{cob1} that the capacity-distortion function is given by
	\begin{align}\label{ex1lu}
		C(D)= \max_{P_{UX}}\big\{ H(X|U)+I(U;Z)\big|\mathbb{E}\big[d\big(S, \hat{S}^*(U,Z)\big)\big]\leq D\big\},
	\end{align}
	which is characterized by a single-parameter form in the following corollary.
	\begin{corollary} \label{co3}
			The capacity-distortion function 
			$C(D)=  {\rm{H}}_2(\alpha)+1-{\rm{H}}_2(\alpha\ast q)$, where $\min \{\alpha\bar{q},\bar{\alpha}q\}+\min \{\bar{\alpha}\bar{q},\alpha q\}=D$ for $\alpha\in [0,1/2]$.
	\end{corollary}
	\begin{proof}		

Without loss of generality, we assume
that $q\geq 1/2$.  Applying Theorem \ref{th1} to Example 1,  we obtain that $R=I(X ; Y | U,S)+I(U;Z)=H(X|U)+H(Z)-H(Z|U)$ and the distortion $\mathbb{E}\big[d\big(S, \hat{S}^*(U,Z)\big)\big]=\sum_{u,z}\min_s p(s,u,z)$. 
Similar to the proof of the capacity region of a binary symmetric broadcast channel (BS-BC), we get that  $R= {\rm{H}}_2(\alpha)+1-{\rm{H}}_2(\alpha\ast q)$ is achievable by applying the superposition coding and successive cancellation decoding, i.e., $X=U\oplus V$, where $U$ and $V$ are independent, $U$ is Bernoulli$(\beta)$, and $V$ is Bernoulli$(\alpha)$ for $\beta \in [0,1]$.   Correspondingly, the distortion $D=\min \{\alpha\bar{q},\bar{\alpha}q\}+\min \{\bar{\alpha}\bar{q},\alpha q\}$. 

In the following, we prove the converse. By leveraging the fact that there exists an $\alpha\in [0,1/2]$ such that $0 \leq H(Z)-H(Z|U)\leq 1-{\rm{H}}_2(q)=1-{\rm{H}}_2(\alpha\ast q)$ and the Mrs. Gerber’s Lemma in \cite{el2011network}, we get that ${\rm{H}}_2(\alpha\ast q)=H(Z|U)\geq {\rm{H}}_2\big({\rm{H}}_2^{-1}\big({\rm{H}}_2(X|U)\big)\ast q\big)$. According to the monotonicity of binary entropy function and combining the facts that $\alpha\ast q \geq 1/2$ and $\alpha\leq e$ if $\alpha\ast q \geq e\ast q$, we have $H(X|U)\leq {\rm{H}}_2(\alpha)$ for $\alpha\in [0,1/2]$, which leads to $R\leq {\rm{H}}_2(\alpha)+1-{\rm{H}}_2(\alpha\ast q)$.  
Furthermore,  since $\left|\mathcal{U}\right| \leq|\mathcal{X}|+1=3$, we set $P(U=i)=\beta_i,i=0,1,2$, where $\sum_{i=0}^{2}\beta_i=1$. Thus, we have $D \geq D_{\min}(\alpha_0,\alpha_1,\alpha_2)=\sum_{i=0}^{2}\beta_i\big(\min \{\alpha_i\bar{q},\bar{\alpha}_iq\}+\min \{\bar{\alpha}_i\bar{q},\alpha_i q\}\big)$, where $p(x=0|u=i)=\alpha_i,i=0,1,2$. From the above expression about $D_{\min}$, we further get that $D_{\min}(\alpha_0,\alpha_1,\alpha_2)\geq \min_{\alpha_i,i=0,1,2}\big\{\min \{\alpha_i\bar{q},\bar{\alpha}_iq\}+\min \{\bar{\alpha}_i\bar{q},\alpha_i q\}\big\}$.
Therefore, there exists an $\alpha\in [0,1/2]$ such that $D_{\min}(\alpha_0,\alpha_1,\alpha_2)\geq D_{\min}(\alpha)=\min \{\alpha\bar{q},\bar{\alpha}q\}+\min \{\bar{\alpha}\bar{q},\alpha q\}$, which completes the proof. 
\end{proof} 
Note that if we continue with the channel model from Example 1 and consider the bistatic ISAC broadcast channel model, then its capacity-distortion region is 
	\begin{equation}
	C_{BC}= \bigcup_{\alpha\in [0,1/2]}\left\{
	\begin{array}{lc}
		(R_1,R_2,D) \in \mathbb{R}_+^3:&  \\
		R_2\leq 1-{\rm{H}}_2(\alpha\ast q) &  \\
		R_1+R_2\leq {\rm{H}}_2(\alpha)+1-{\rm{H}}_2(\alpha\ast q) & \\
		D\geq\min \{\alpha\bar{q},\bar{\alpha}q\}+\min \{\bar{\alpha}\bar{q},\alpha q\} 
	\end{array}
	\right\}.
\end{equation}	

For comparison, we also derive a lower bound and an upper bound on the capacity-distortion function based on Corollary \ref{co33} and Theorem \ref{tho}, respectively. Applying Corollary \ref{co33} to Example 1, we obtain the following results.
\begin{corollary} \label{co4}
	If $D=\min \{p\bar{q},\bar{p}q\}+\min \{\bar{p}\bar{q},pq\}$, then $({\rm{H}}_2(p),D)$ is achievable for $p\in [0,1]$.			
\end{corollary}
\begin{proof}
	According to the lower bound given in Corollary \ref{co33}, we have that
	$I(X;Y|S)=H(X)= {\rm{H}}_2(p)$ and $D=E[d(S^*(Z),S)]=\sum_{z}p(z)\sum_{s}p(s|z)d(s^*(z),s)$.
	Furthermore, we obtain that $s^*(z)=\arg\max_{s}p(s|z)$ and $D=\sum_{z}\min p(s,z)=\min \{p\bar{q},\bar{p}q\}+\min \{\bar{p}\bar{q},pq\}$.
\end{proof}
	Further, based on Theorem \ref{tho}, we give a more concise upper bound expression of the capacity-distortion region for this specific channel model as follows.	
	\begin{corollary}\label{thou}
		Let $D$ be the minimum distortion achievable at rate $R$, then $D=0$ for $R\leq 1-{\rm{H}}_2(q)$; $D\ge {\rm{H}}_2^{-1}\big(R-1+{\rm{H}}_2(q)\big)$, for $1-{\rm{H}}_2(q)< R\leq 1$.		
	\end{corollary}
	\begin{proof}
See Appendix \ref{prco5} for details.
	\end{proof}
	\begin{figure}[h]
		\centering
		\includegraphics[width=4in]{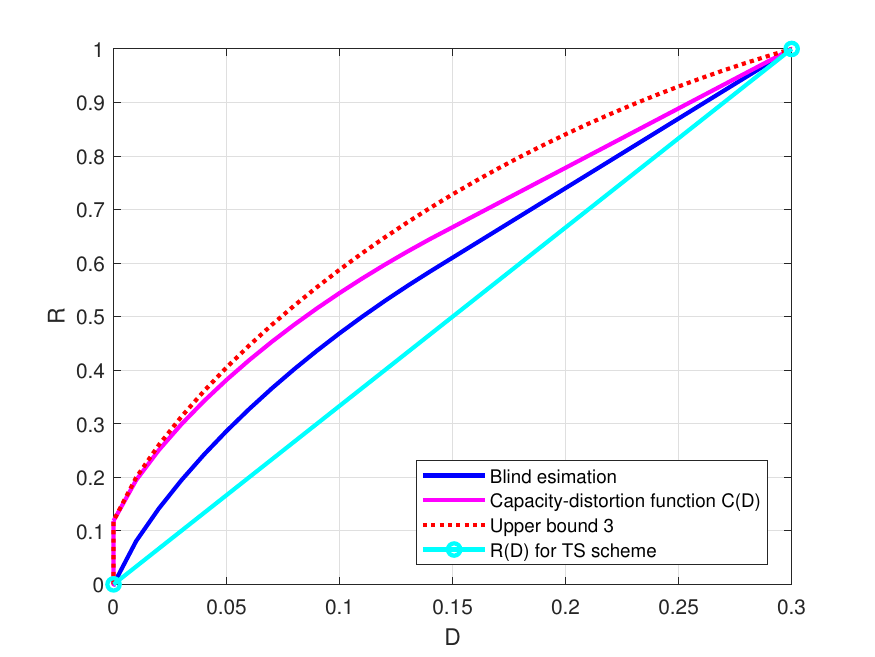}
		\vspace{-6mm}
		\caption{The capacity-distortion functions with parameter $q = 0.7$ in Example 1.}
		\label{exa1}
	\end{figure}
	In the following, we numerically evaluate all the above single-letter bounds on the capacity-distortion functions. As shown in Fig. \ref{exa1}, the capacity-distortion function is achieved by the bound corresponding to the partial-decoding-based estimation strategy, which outperforms that of the blind estimation strategy. The upper bound 3 is distant from the optimal bound to some degree, particularly when the distortion is significant.
	Additionally, we compare the performance of the proposed bounds  with the basic time-sharing scheme (TS). The TS scheme refers to independent communication and sensing in a time-sharing manner. In the TS scheme, the minimum distortion $D_{\min }=0$ is achieved always by sending $X=1$, i.e., $P(x=1)=1$ and $P(x=0)=0$. Furthermore, we obtain $R_{\min}=0$, $R_{\max}=1$, and $D_{\max}=\min \{q, 1-q\}$ by using the best constant estimator, $\hat{s}=\operatorname{argmax}_{s \in\{0,1\}} P_S(s)$. The TS scheme thus achieves the straight line between $(0,0)$ and $(0.3, 1)$. 
	We observe a gain of the proposed bounds over the baseline TS scheme, which indicates the benefits of exploiting communications to assist sensing for a bistatic system.
	
	%
	%
\subsection{Example 2}
		We consider a binary communication channel with a multiplicative Bernoulli state, where the signal received by the ComRx is given by $Y=S X$ with binary alphabets $\mathcal{X}=\mathcal{S}=\mathcal{Y}=\{0,1\}$ and the state $S$ is Bernoulli$(q)$, for $q \in(0,1)$. The signal received by the SenRx is expressed as the sum of two parts, where one part is the signal reflected by the ComRx and the other part is the directly radiated signal by of the ISAC Tx. The first part is expressed as the product of a binary channel with a multiplicative Bernoulli state and an attenuation factor $a \in (0,1)$ and the second part is expressed as a binary symmetric channel $\big(\mathrm{BSC}(e)\big)$ with error probability $e \in [0,1]$. Hence, the sensing channel is expressed as  $Z=a S X+(X\oplus N) $ with alphabets  $\mathcal{Z}=\{0,1,a,a+1\}$, where $a \in (0,1)$, $N$ is Bernoulli$(e)$, and $e \in [0,1]$. We still consider the Hamming distortion measure $d(s, \hat{s})=s \oplus \hat{s}$ to characterize the sensing accuracy.  

\begin{figure}[h]
		\centering
		\includegraphics[width=4in]{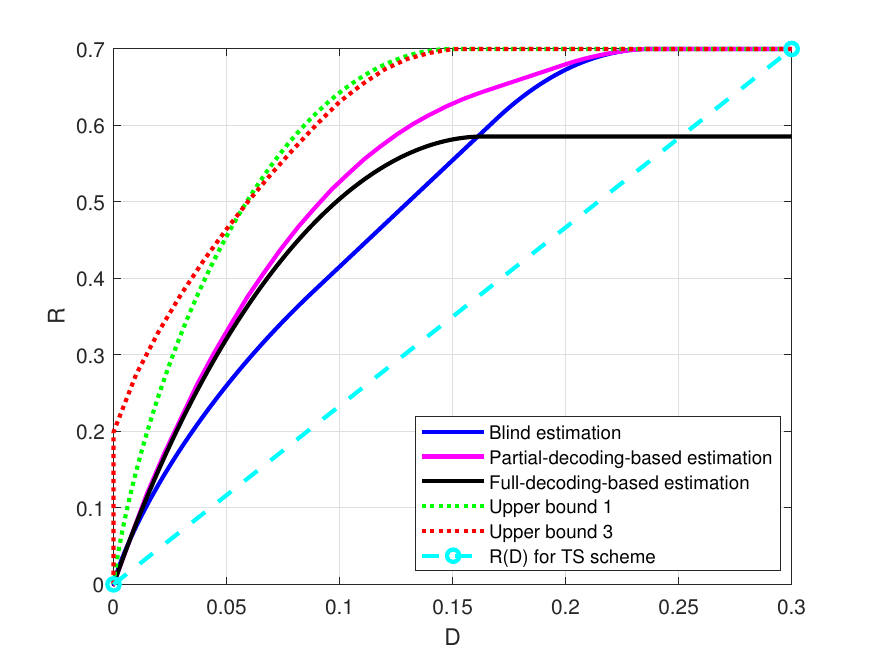}
		\caption{The capacity-distortion functions with parameters $e = 0.25$ and $q = 0.7$ in Example 2.}
		\label{5}
	\end{figure}

	Applying the theoretical results in Section \ref{secp} to Example 2, we obtain the rate-distortion functions in Appendix \ref{prco4} and the corresponding curves as shown in Fig. \ref{5}.
	
As shown in Fig. \ref{5}, the lower bound corresponding to the partial-decoding-based estimation strategy outperforms those of both the blind estimation and full-decoding-based estimation strategies, demonstrating that the partial-decoding-based estimation strategy strikes a better balance between sensing and communication. Furthermore, the two upper bounds in the upper left of the curves for the three DnE strategies exhibit similar performance. Additionally, we observe a significant improvement in the proposed lower bounds over a baseline TS scheme. The performance gap underscores the benefits of coding design and the advantage of leveraging communication to assist sensing in a bistatic system.

	\section{ Conclusion}\label{secco}
	 In this paper, we considered a bistatic ISAC system in which the sensing receiver is unaware of the sent message. 
	Under this model, we provided a multi-letter characterization of the capacity-distortion function and transformed the resulting infinite-dimensional optimization problem into an infinite sequence of finite-dimensional optimization problems.	
	For ease of calculation, we derive three single-letter lower bounds based on the three \emph{decoding-and-estimation} (DnE) strategies: blind estimation,  partial-decoding-based estimation, and full-decoding-based estimation, respectively. Additionally, three single-letter upper bounds are proposed by introducing the genie estimators and the lossy source coding. In particular, for the special case where the channel to SenRx is  degraded with respect to the channel to ComRx, we provide the single-letter characterization of the capacity-distortion function,  demonstrating the optimality of the partial-decoding-based estimation strategy achieved by superposition coding.  Furthermore,  we extended these results to a bistatic ISAC degraded broadcast channel model, deriving the corresponding capacity-distortion region. Finally, we applied these theoretical results to two examples, illustrating the significance of the proposed bounds and  the benefits of leveraging communications to assist sensing in a bistatic system.

	\appendices
	\section{Proof of Theorem \ref{th2}} \label{prth}
	The proof is  divided into two parts as follows.
	
		1) $C_{S,E}(D) \leq C(D)$
	
	Let a stationary and ergodic stochastic process $X'$ achieves $C_{S,E}(D)-\epsilon$,  where the corresponding distortion satisfies 
	$
		\lim_{k \rightarrow \infty}\frac{1}{k} \mathbb{E}\big[d\big(S^k, \hat{S}^{*k}(Z^k)\big)\big]= D
	$
	according to the definition of $C_{S,E}(D)$ and Lemma \ref{l1}.
	Then, given $\epsilon>0$, for sufficiently large $k$,  $\frac{1}{k} \mathbb{E}\big[d\big(S^k, \hat{S}^{*k}(Z^k)\big)\big]\leq D+\epsilon$. Furthermore, since $I(X' ; Y' | S')+\epsilon= \lim_{k \rightarrow \infty} \frac{1}{k}  I(X^k ; Y^k | S^k),$ there exists a $p(x^k)$ such that $ \frac{1}{k}  I(X^k ; Y^k | S^k)\geq I(X' ; Y' | S')-\epsilon$ for sufficiently large $k$. Therefore, we conclude that 
	$C(D+\epsilon)=\lim_{k \to \infty}\sup_{p(x^k)}\frac{1}{k}  I(X^k ; Y^k | S^k)\geq I(X' ; Y' | S')-\epsilon=C_{S,E}(D)-\epsilon$.
	Taking $\epsilon\to 0$ and using Lemma \ref{l1}, we get 
	$
	C(D)\geq C_{S,E}(D).
	$

	2) $C_{S,E}(D) \geq C(D)$

	Fix $k$ and a pmf $ p(x^k) $ satisfying $ \mathbb{E}\big[d\big(S^k, \hat{S}^{*k}(Z^k)\big)\big]\leq kD $ and $C(D)-\epsilon \leq\frac{1}{k}  I(X^k ; Y^k | S^k)\leq C(D)$. 
	Define  a process  $\{X_i\}_{i=-\infty}^{\infty}$ as $X_{jk+1}^{(j+1)k}$  with $p(x_{jk+1}^{(j+1)k})=p(x^k)$ for all $j$. For $j_1\neq j_2$, $X_{j_1k+1}^{(j_1+1)k}$ and $X_{j_2k+1}^{(j_2+1)k}$ are independent and  $\{S_i\}_{i=-\infty}^{\infty}$ can be defined similarly. Let $Y_i=X_i+S_i$ be the corresponding output process. 
	For any code length $n>k$ and $n=kt+r,r<k$, we have 
	\begin{align}
		I(X^n;Y^n|S^n)
		&=H(X^n)-H(X^n|Y^n,S^n)\notag \\
		&\geq H(X^{tk})+H(X_{tk+1}^{n})-H(X^{tk}|Y^{tk},S^{tk})- H(X_{tk+1}^{n}|Y_{tk+1}^{n},S_{tk+1}^{n})\notag \\
		&= I(X^{tk};Y^{tk}|S^{tk})+I(X_{tk+1}^{n};Y_{tk+1}^{n}|S_{tk+l}^n)\notag \\
		&\geq I(X^{tk};Y^{tk}|S^{tk})\notag \\
		&\geq (tk+r-k) \frac{1}{k}I(X^{k};Y^{k}|S^{k})+(k-r) \frac{1}{k}I(X^{k};Y^{k}|S^{k}).
	\end{align}
	Thus, we get that 
	\begin{align*}
		\frac{1}{n}I(X^n;Y^n|S^n)+\epsilon_m\geq \frac{1}{k}I(X^{k};Y^{k}|S^{k}),
	\end{align*}
	where $\epsilon_m= \frac{1}{n}I(X^{k};Y^{k}|S^{k})>0$. 
	
	For each $t=0,1,\dots,k-1$, define the $\{X_i(t)\}_{i=-\infty}^{\infty}$ as $X_{t+i}$ and similarly define $\{S_i(t)\}_{i=-\infty}^{\infty}$ and $\{Y_i(t)\}_{i=-\infty}^{\infty}$. Constructing a  stochastic process $\{\tilde{X}_i\}_{i=-\infty}^{\infty} $ with the joint probability distribution $p(\tilde{X}_j
	^m=x_j^m)=\frac{1}{k}\sum_{t=0}^{k-1} p(X_j^m(t)=x_j^m)$, and similarly define $\{\tilde{S}_i\}_{i=-\infty}^{\infty} $ and $\{\tilde{Y}_i\}_{i=-\infty}^{\infty} $.  According to the definition of stochastic process $\{\tilde{X}_i\}_{i=-\infty}^{\infty}$, we have 
	\begin{align*}
		&p(\tilde{X}_{j+1} 
		^{m+1}=x_j^m )=\frac{1}{k}\sum_{t=0}^{k-1} p\big(X_{j+1}^{m+1}(t)=x_j^m \big)\\
		=&\frac{1}{k}\sum_{t=0}^{k-1} p\big(X_{j}^{m}(t)=x_j^m \big)=p(\tilde{X}_j
		^m=x_j^m)
	\end{align*}
	due to $p(X_{j+k}^{m+k})=p(X_{j}^{m})$. Thus, $\{\tilde{X}_i\}_{i=-\infty}^{\infty} $ is a stationary stochastic process.  The ergodicity of $\{\tilde{X}_i\}_{i=-\infty}^{\infty}$ can be obtained by referring to \cite[Lemma 1]{feinstein1959coding}.
	
	In order to more clearly express the connection between $\{\tilde{X}_i\}_{i=-\infty}^{\infty} $ and $\{X_i(t)\}_{i=-\infty}^{\infty} $, we introduce
	a random variable $T$ that is independent of everything else and uniformly distributed over $\{0,1,\dots,k-1\}$.  Observe that
	\begin{align*}
	 p\big(X_i(T)\big)=\sum_{t=0}^{k-1}p(T=t)p\big(X_i(T)|T=t\big)=\frac{1}{k}\sum_{t=0}^{k-1} p(X_i(t)\big)=\frac{1}{k}\sum_{t=0}^{k-1} p\big(X_{i+t}\big),
	\end{align*}
	  for all $i$, thus, we use $\{X_i(T)\}_{i=-\infty}^{\infty}$ instead of $\{\tilde{X}_i\}_{i=-\infty}^{\infty} $ in the proof below. Similarly, we can define $p\big(S_i(T)\big)$ and $p\big(Y_i(T)\big)$.
	
	In the following, we prove that the stationary and ergodic stochastic process $\{\tilde{X}_i\}_{i=-\infty}^{\infty} $ satisfies the distortion constraint. Considering the definition of distortion, we have
	\begin{align}\label{eds}
		&\mathbb{E}\Big[d\Big(S^n(T), \hat{S}^{*n}\big(Z^n(T)\big)\Big)\Big]\notag\\
		=&\mathbb{E}\bigg[\mathbb{E}\Big[d\Big(S^n(T), \hat{S}^{*n}\big(Z^n(T)\big)\Big)\Big]\bigg| T\bigg]\notag\\
		=&\frac{1}{k}\sum_{i=0}^{k-1}\mathbb{E}\Big[d\Big(S^n(i), \hat{S}^{*n}\big(Z^n(i)\big)\Big)\Big].
	\end{align} 
	Then, we state that $\lim_{n \rightarrow \infty}\frac{1}{n} \mathbb{E}\Big[d\Big(S^n(j), \hat{S}^{*n}\big(Z^n(j)\big)\Big)\Big]\leq D$, for every $0 \leq j \leq k-1$. Considering $\mathbb{E}\Big[d\Big(S^n(j), \hat{S}^{*n}\big(Z^n(j)\big)\Big)\Big]$ for some $0 \leq j \leq k-1$, we have
	\begin{align}\label{ned}
		&\mathbb{E}\Big[d\Big(S^n(j), \hat{S}^{*n}\big(Z^n(j)\big)\Big)\Big]
		=\mathbb{E}\big[d\big(S_{1+j}^{n+j}, \hat{S}_{1+j}^{*(n+j)}(Z_{1+j}^{n+j})\big)\big]
		=\sum_{i=1+j}^{n+j}\mathbb{E}\big[d\big(S_i, \hat{S}_i^{*}(Z_{1+j}^{n+j})\big)\big]\notag\\
		=&\sum_{i=1+j}^{k}\mathbb{E}\big[d\big(S_i, \hat{S}_i^{*}(Z_{1+j}^{n+j})\big)\big]+\sum_{i=k+1}^{2k}\mathbb{E}\big[d\big(S_i, \hat{S}_i^{*}(Z_{1+j}^{n+j})\big)\big]+\notag
		\cdots \\ &\quad +\sum_{i=(t-1)k+1}^{tk}\mathbb{E}\big[d\big(S_i, \hat{S}_i^{*}(Z_{1+j}^{n+j})\big)\big]+\sum_{i=tk+1}^{n+j}\mathbb{E}\big[d\big(S_i, \hat{S}_i^{*}(Z_{1+j}^{n+j})\big)\big]\notag\\
		=&\sum_{i=1+j}^{k}\mathbb{E}\big[d\big(S_i, \hat{S}_i^{*}(Z_{1+j}^{n+j})\big)\big]+\sum_{i=k+1}^{2k}\mathbb{E}\big[d\big(S_i, \hat{S}_i^{*}(Z_{k+1}^{2k})\big)\big]+\notag
		\cdots \\ &\quad +\sum_{i=(t-1)k+1}^{tk}\mathbb{E}\big[d\big(S_i, \hat{S}_i^{*}(Z_{(t-1)k+1}^{tk})\big)\big]+\sum_{i=tk+1}^{n+j}\mathbb{E}\big[d\big(S_i, \hat{S}_i^{*}(Z_{1+j}^{n+j})\big)\big]\notag\\
		\leq & (k-j)d_{\max}+(t-1)kD+(r+j)d_{\max}\notag\\
		=&(t-1)kD+(k+r)d_{\max}
	\end{align}		
	From (\ref{ned}) we get that $\lim_{n \rightarrow \infty}\frac{1}{n} \mathbb{E}\Big[d\Big(S^n(j), \hat{S}^{*n}\big(Z^n(j)\big)\Big)\Big]\leq D$. Thus, for code length $n$, we obtain that the distortion  satisfies $\lim_{n \rightarrow \infty}\frac{1}{n}\mathbb{E}\big[d\big(S^n, \hat{S}^{*n}(Z^n)\big)\big]\leq D$ .
	
	In the end, we prove that $	\lim_{n \rightarrow \infty}I(\tilde{X}^n;\tilde{Y}^n|\tilde{S}^n)\geq \frac{1}{k}I(X^{k};Y^{k}|S^{k})$.
	According to \cite[Lemma 2, Lemma 3]{feinstein1959coding}, we get the above stationary construction way does not change the entropy limit, which means
	\begin{align*}
		\lim_{n \rightarrow \infty}I(\tilde{X}^n;\tilde{Y}^n|\tilde{S}^n)=\lim_{n \rightarrow \infty}I(X^n;Y^n|S^n).
	\end{align*}
	Thus, we get that 
	\begin{align*}
		I(X' ; Y' | S')= \lim_{n \rightarrow \infty}I(\tilde{X}^n;\tilde{Y}^n|\tilde{S}^n)=\lim_{n \rightarrow \infty}I(X^n;Y^n|S^n)\geq\frac{1}{k}I(X^{k};Y^{k}|S^{k}).
	\end{align*}
	Furthermore, we get  $C_{S,E}(D) \geq C(D)$ according to the definitions of $C_{S,E}(D)$ and $C(D)$.

	\section{Proof of the optimality of the one-shot state estimator} \label{prl1}
	
	According to (\ref{D}) and the law of total expectation, we have
	\begin{align}
	\quad \ \mathbb{E}\big[d\big(S_i, \hat{S}_i(U_i,Z_i)\big)\big|U_i=u_i\big]
		&= \sum_{z_i,\hat{s},s} P_{Z_i|U_i}P_{\hat{S}_i |U_iZ_i} P_{S_i|U_iZ_i} d(s, \hat{s}) \notag\\
		&\geq \sum_{z_i} P_{Z_i|U_i}\min _{\hat{s} \in \mathcal{S}} \sum_s P_{S_i| U_iZ_i} d(s, \hat{s})\notag \\
		&\stackrel{(\mathrm{a})}{=}  \sum_{z_i,s} P_{Z_i|U_i}P_{S_i| U_iZ_i}d\big(s, \hat{s}^*(u_i,z_i)\big) \notag\\
		&= \mathbb{E}\big[d\big(S_i, \hat{s}^*(U_i,Z_i)\big)\big|U_i=u_i\big],\notag
	\end{align}
where $(a)$ holds by choosing $\hat{s}^*(u, z)$ in Theorem \ref{th1}. This yields the desired conclusion.	

\section{Proof of Theorem \ref{tho2}} \label{prth5}		
	According to Fano's inequality \cite{el2011network}, we have		
	\begin{align}\label{outer1n1}
		&\quad \ n R-n \epsilon_n\leq I\left(W; Y^n,S^n\right)= I\left(W ; Y^n|S^n\right)\notag\\
		&= \sum_{i=1}^n I\left(W;Y_i|Y^{i-1},S^n\right)= \sum_{i=1}^n H\left(Y_i|Y^{i-1},S^n\right)-H\left(Y_i|X_i,W,Y^{i-1},S^n\right)\notag\\
		&\stackrel{(a)}{\leq}  \sum_{i=1}^n H\left(Y_i|S_i\right)-H\left(Y_i|X_i,S_i\right) =  \sum_{i=1}^nI(X_i;Y_i|S_i),
	\end{align}
	where $(a)$ holds by the Markov chain $\left(W, Y^{i-1}, S^{i-1},S_{i+1}^n \right) - (X_i,S_i) - Y_i$. 	
	\begin{align}\label{outer1n2}
		& \quad \ n R-n \epsilon_n\leq I\left(W ; Y^n,S^n\right) \notag\\
		&\leq \sum_{i=1}^nI\left(W,Z_{i+1}^n ; Y_i,S_i|Y^{i-1},S^{i-1}\right)+I\left(Z_{i+1}^n ; Z_i\right)\notag\\
		&= \sum_{i=1}^nI\left(Z_{i+1}^n ; Y_i,S_i|Y^{i-1},S^{i-1}\right)+I\left(W; Y_i,S_i|Y^{i-1},S^{i-1},Z_{i+1}^n\right)+I\left(Z_{i+1}^n ; Z_i\right)\notag\\
		&\stackrel{(a)}{=} \sum_{i=1}^nI\left(Y^{i-1},S^{i-1} ; Z_i|Z_{i+1}^n\right)+I\left(W; Y_i,S_i|Y^{i-1},S^{i-1},Z_{i+1}^n\right)+I\left(Z_{i+1}^n ; Z_i\right)\notag\\
		&= \sum_{i=1}^n I\left(X_i; Y_i,S_i|U_i \right) + I\left(U_i; Z_i\right)\stackrel{(b)}{=} \sum_{i=1}^n I\left(X_i; Y_i|S_i,U_i \right) + I\left(U_i; Z_i\right)		
	\end{align}
	where $Y^0,S^0, Z_{n+1}^n=\emptyset$, $(a)$ follows by the Csiszar sum identity \cite{el2011network}, the auxiliary random variable identification $U_i=\left(Z_{i+1}^n,Y^{i-1},S^{i-1}\right)$ and $(b)$ holds by the independence of $(U,X)$ and $S$. 
	
	Furthermore, replacing $Y$ in the above derivation with $(Y,Z)$ and introducing $V_i=\left(Z^{i-1}\right)$,  we get
	\begin{align}\label{outer1n3}
		& \quad \ n R-n \epsilon_n\notag\\
		&\leq \sum_{i=1}^nI\left(Y^{i-1},Z^{i-1},S^{i-1} ; Z_i|Z_{i+1}^n\right)+I\left(X; Y_i,Z_i,S_i|Y^{i-1},Z^{i-1},S^{i-1},Z_{i+1}^n\right)+I\left(Z_{i+1}^n ; Z_i\right)\notag\\
		&= \sum_{i=1}^n I\left(X_i; Y_i,Z_i,S_i|U_i,V_i \right) + I\left(U_i,V_i; Z_i\right)\notag\\
		&= \sum_{i=1}^n I\left(X_i; Y_i,Z_i|S_i,U_i,V_i \right) + I\left(U_i,V_i; Z_i\right)		
	\end{align}
	The rest of the proof of the rate expression follows by exploiting the time-sharing random variable $Q \sim$ Unif $[1: n]$ introduced in the proof of Theorem \ref{tho2}. The cardinality bounds on $U$ and $V$ can be proved using the extension of the convex cover method to multiple random variables in \cite[Appendix C]{el2011network}.

	\section{Proof of Theorem \ref{th3}} \label{prth7}
	The proof is  divided into an achievability part and a converse part.
	
		{\bf Proof of achievability:} Applying superposition coding, with the idea of rate splitting, we divide the private message $W_1$ into two independent messages $W_{10}$ at rate $R_{10}$ and $W_{11}$ at rate $R_{11}$, as in the proof of \cite[Theorem 8.1]{el2011network}. Let the cloud center $U$ in the superposition coding represents the message pair $(W_2,W_{10})$ and the satellite codeword $X$ represents the message triple $(W_2,W_{10},W_{11})$. Following similar steps to proof of Theorem \ref{th1} and eliminating $R_{10}$ by the Fourier–Motzkin procedure, we can show that $(R_1,R_2,D)$ is achievable if 
	\begin{align*}
		&R_2\leq I(U;Z), R_1+R_2\leq I(X ; Y | U,S)+I(U;Z), D \geq \mathbb{E}[d\big(S, \hat{s}^*(U,Z)\big)\big],
	\end{align*}	
	which completes the proof of the achievability. 
	
	{\bf Proof of converse:} Using Fano's inequality \cite{el2011network} and the property of the degraded channel, we obtain 		
	\begin{align}\label{outer11}
		& \quad \ n \left(R_1 + R_2\right)-n \epsilon_n\leq I\left(W_1 ; Y^n,S^n|W_0\right)+I\left(W_0 ; Z^n\right) \notag\\
		& \stackrel{(a)}{=} \sum_{i=1}^nI\left(W_1 ; Y_i,S_i|W_0,Y^{i-1},S^{i-1}\right) + I\left(W_0 ; Z_i|Z_{i+1}^n\right)\notag\\
		& \leq \sum_{i=1}^n I\left(W_1,Z_{i+1}^n ; Y_i,S_i|W_0,Y^{i-1},S^{i-1}\right) + I\left(W_0 ,Z_{i+1}^n; Z_i\right)\notag\\
		&\stackrel{(b)}{=} \sum_{i=1}^n I\left(W_1; Y_i,S_i|W_0,Y^{i-1},S^{i-1},Z_{i+1}^n \right) + I\left(W_0 ,Z_{i+1}^n,Y^{i-1},S^{i-1};Z_i\right)\notag\\
		&\leq \sum_{i=1}^n I\left(X_i; Y_i,S_i|W_0,Y^{i-1},S^{i-1},Z_{i+1}^n \right) + I\left(W_0 ,Z_{i+1}^n,Y^{i-1},S^{i-1};Z_i\right)\notag\\
		&\stackrel{(c)}{=} \sum_{i=1}^n I\left(X_i; Y_i|S_i,W_0,Z_{i+1}^n,Y^{i-1},S^{i-1},Z^{i-1} \right) + I\left(W_0 ,Z_{i+1}^n,Y^{i-1},S^{i-1},Z^{i-1};Z_i\right)\notag\\
		&= \sum_{i=1}^n I\left(X_i; Y_i|S_i,U_i \right) + I\left(U_i; Z_i\right),
	\end{align}
	where $Y^0=S^0=Z_{n+1}^n=\emptyset$, $(a)$ follows from the chain rule for mutual information, $(b)$ follows from the Csiszar sum identity \cite{el2011network}, the auxiliary random variable identification  $U_i=\left(W_0,Z_{i+1}^n,Y^{i-1},S^{i-1},Z^{i-1}\right)$, and $(c)$ follows by the Markov chain  $Y_i - Y^{i-1} - Z^{i-1}$ given $(S_i,W_0,Z_{i+1}^n,S^{i-1})$, 
	the Markov chain $Z_{i} - (W_0,Y^{i-1}) - Z^{i-1}$, and the independence of $(U,X)$ and $S$.
	
	Similarly, we have
	\begin{align}\label{outer12}
		\quad \ n R_0 -n \epsilon_n\leq I(W_0 ; Z^n)\leq \sum_{i=1}^n I(W_0 ,Z_{i+1}^n,Y^{i-1},S^{i-1},Z^{i-1}; Z_i)= \sum_{i=1}^n I\left(U_i; Z_i\right).
	\end{align}	
	The rest of proof of the rate expression follows by introducing a time-sharing random variable $Q \sim$ Unif $[1: n]$ independent of $\left(W_0, W_1, X^n, Y^n, Z^n, S^n\right)$ and by defining $U=\big(Q, \tilde{U}_Q\big), X=X_Q, Y=Y_Q$, $Z=Z_Q$ and $S=S_Q$. 	
	
The proof of the distortion constraint is identical to that in Theorem \ref{tho2} and is omitted here.
	
	\section{Proof of Corollary \ref{thou}} \label{prco5}
	
	For the case that $R\le 1-{\rm{H}}_2(q)$, let $\mathcal{C}$ be a code with rate $R$. According to the channel model $Y=S\oplus X$ and the fact that the ComRx is accessible to the channel state, the communication channel to ComRx is perfect. Thus, the ComRx can always recover the message correctly. For the SenRx, since $R\le 1-{\rm{H}}_2(q)$, SenRx can recover the codeword $X^n$ through decoding with high probability, thus the minimum distortion achievable at rate $R$ is $D=0$. 
	
	For the case that $1\geq R>1-{\rm{H}}_2(q)$. Let $\mathcal{C}^n$ be a code with rate $R$ and length $n$ and $D$ be distortion corresponding to $\mathcal{C}^n$. Denote the codeword sent through the channel as $X^n$, the output of the channel as $(Y^n, Z^n)$, and the estimate of channel state $S^n$ as $\hat{S}^n$. Define that $\hat{X}_i=Z_i\oplus \hat{S}_i$ and  $P(\hat{X}_i \ne X_i)=p_i$. Then we get
	$$\mathbb{E}[d(\hat{S}_i,S_i)]=P(\hat{S}_i \ne S_i)=P(\hat{X}_i \ne X_i)=p_i$$ and 
	$$
	\sum_{i=1}^{n}\mathbb{E}[d(\hat{S}_i,S_i)]\le nD.
	$$ Using Fano's inequality, we have that
	\begin{align}\label{fa} 
		&\frac{1}{n}H(X^n)-\frac{1}{n}I(X^n;Z^n)
		\leq  \frac{1}{n}H(X^n)-\frac{1}{n}I(X^n;\hat{X}^n)\notag\\
		\leq &\frac{1}{n}\sum_{i=1}^{n}H(X_i|\hat{X}^n)
		\leq  \frac{1}{n}\sum_{i=1}^{n}H(X_i|\hat{X}_i)
		\leq \frac{1}{n}\sum_{i=1}^{n}{\rm{H}}_2(p_i)+p_i\log(|\mathcal{X}|-1)\notag\\
		\leq & {\rm{H}}_2\bigg(\frac{1}{n}\sum_{i=1}^{n} p_i\bigg)
		= {\rm{H}}_2\bigg(\frac{1}{n}\sum_{i=1}^{n}\mathbb{E}[d(\hat{S}_i,S_i)]\bigg).
	\end{align}
	Hence we obtain
	\begin{align}\label{fa1} 
		\frac{1}{n}H(X^n)-\frac{1}{n}I(X^n;Z^n)
		\leq  {\rm{H}}_2\bigg(\frac{1}{n}\sum_{i=1}^{n}\mathbb{E}[d(\hat{S}_i,S_i)]\bigg).
	\end{align}
	Combining the fact that $H(X^n)=nR$ with  $\frac{1}{n}I(X^n;Z^n)\le 1-{\rm{H}}_2(q)$, we get ${\rm{H}}_2^{-1}\big(R-1+{\rm{H}}_2(q)\big)\le D$, which completes the proof.

	
	\section{Bounds on the capacity-distortion function of Example 2} \label{prco4}
	
	\subsection {Three lower bounds to Example 2} \label{inne2}
	Throughout this section  let $p\triangleq P_X(0)$. Applying Corollary \ref{co33} to Example 2, since $Y$ is deterministic given $(S, X)$ and it equals 0 whenever $S=0$, we have 
	\begin{align}
		I(X ; Y | S) & =  p_S(0) H(Y | S=0)+p_S(1) H(Y | S=1) =  p_S(1) H(X)=q {\rm{H}}_2(p)\notag .
	\end{align}
	Notice that the estimator at the SenRx accurately estimate the state $\hat{S}=S=1 $ when $z=a$ or $z=a+1$. Since the distortion occurs at $z=0$ and $z=1$, we then calculate the distortion in two cases. For $z=0$, we need to compare $p(s=0|z=0)$ and $p(s=1|z=0)$ and select the corresponding $S$ with a higher probability as the estimate of $S$. And we calculate that
	\begin{align}\label{s0z0}
		p(s=0|z=0)=\frac{p(s=0,z=0)}{p(z=0)} =\frac{p(s=0)p(z=0|s=0)}{p(z=0)}=\frac{\bar{q}(p\bar{e}+\bar{p}e)}{p(z=0)}	
	\end{align}
	\text{and}
	\begin{align}
		p(s=1|z=0)&=\frac{qp\bar{e}}{p(z=0)}.\label{s1z0}
	\end{align}
	Similarly, for $z=1$, we have 
	\begin{align}
		p(s=0|z=1)&=\frac{\bar{q}(pe+\bar{p}\bar{e})}{p(z=1)}\label{s0z1}	
	\end{align} 
	\text{and}
	\begin{align}
		p(s=1|z=1)&=\frac{qpe}{p(z=1)}.\label{s1z1}
	\end{align}
	As a result, we get that $\hat{S}=0$ if $(\ref{s0z0})\geq (\ref{s1z0})$ and $\hat{S}=1$, otherwise. Then, we have $\hat{S}=0$ if $(\ref{s0z1})\geq (\ref{s1z1})$ and $\hat{S}=1$, otherwise. 
	
	Therefore, we obtain
	\begin{align*}
		C(D)\ge \max_{p\in[0,1]}\{q{\rm{H}}_2(p):\ d_{11}+d_{12}\le D\}
	\end{align*}	
	where 
	\begin{align}
		d_{11}=\left\{\begin{matrix}
			\bar{q}(\bar{p}e+p\bar{e}), &   qp\bar{e}/(p\bar{e}+\bar{q}\bar{p}e)\geq  1/2,\\
			qp\bar{e} , &   \text{else},
		\end{matrix}\right.
	\end{align}
	\text{and} 
	\begin{align}
		d_{12}=\left\{\begin{matrix}
			\bar{q}(\bar{p}\bar{e}+pe) , &  qpe/(pe+\bar{q}\bar{p}\bar{e}) \geq 1/2 ,\\
			qpe, &   \text{else}.
		\end{matrix}\right.
	\end{align}	
	
	Applying Theorem \ref{th1} to Example 2, since $\left|\mathcal{U}\right| \leq|\mathcal{X}|+1$, without loss of generality, we take $\mathcal{U}=\{0,1,2\}$, $u\in \mathcal{U}$, and let the joint distribution of $U$ and $X$ be $p_{jk}=p(u=j,x=k),j=0,1,2, k=0,1$ and define $p_{uj}=p(u=j),j=0,1,2$. Next, we have
	\begin{align*}
		I\left(X;Y|U,S\right)=  qH(Y|U,S=1)
		= q\Big(p_{u0}{\rm{H}}_2\big(\frac{p_{00}}{p_{u0}}\big)+p_{u1}{\rm{H}}_2\big(\frac{p_{10}}{p_{u1}}+p_{u2}{\rm{H}}_2\big(\frac{p_{20}}{p_{u2}}\big)\Big).
	\end{align*} 
	
	According to $\hat{s}^*(u, z)$ in Theorem \ref{th1}, for $z=0$, we calculate that
	\begin{align*}
		&p(s=0|z=0,u=0)=\frac{p(s=0,z=0,u=0)}{p(u=0,z=0)} \\
		=&\frac{\sum_{k=0}^{1} p(z=0|s=0,x=k)p(s=0,u=0,x=k)}{p(u=0,z=0)}\\
		=& 1-\frac{q\bar{e}p_{00}}{\bar{e}p_{00}+\bar{q}ep_{01}}
	\end{align*}
	\text{and}
	\begin{align*}
		p(s=1|z=0,u=0)=\frac{q\bar{e}p_{00}}{\bar{e}p_{00}+\bar{q}ep_{01}}.
	\end{align*}
	Thus, for $j=0,1,2$, we have
	\begin{align}
		p(s=0|z=0,u=j)&=1-\frac{q\bar{e}p_{j0}}{\bar{e}p_{j0}+\bar{q}ep_{j1}},\label{z0s0}\\
		p(s=1|z=0,u=j)&=\frac{q\bar{e}p_{j0}}{\bar{e}p_{j0}+\bar{q}ep_{j1}},\label{z0s1}\\
		p(s=0|z=1,u=j)&=1-\frac{qep_{j0}}{ep_{j0}+\bar{q}\bar{e}p_{j1}},\label{z1s0}
	\end{align} 
	\text{and}
	\begin{align}
		p(s=1|z=1,u=j)=\frac{qep_{j0}}{ep_{j0}+\bar{q}\bar{e}p_{j1}}\label{z1s1}.
	\end{align}
	As a result, we get that $\hat{S}=0$ if $(\ref{z0s0})\geq (\ref{z0s1})$ and $\hat{S}=1$, otherwise. Similarly, we have $\hat{S}=0$ if $(\ref{z1s0})\geq (\ref{z1s1})$ and $\hat{S}=1$, otherwise. 
	
	Therefore, 
	we obtain that the capacity-distortion function 
	\begin{align*}
		C(D)&\ge \max_{p_{UX}}\{\min\{q{\rm{H}}_2(p),I\left(U ; Z \right)+I\left(X ; Y | U, S\right)\}:\sum_{i=0}^{1}\sum_{j=0}^{2}p(z=i,u=j)d_{ij}\le D\}
	\end{align*}	
	where 	\begin{equation}\label{d0012}
		\begin{split}
			d_{0j}=\left\{\begin{matrix}
				1-\frac{q\bar{e}p_{j0}}{\bar{e}p_{j0}+\bar{q}ep_{j1}},  &   q\bar{e}p_{j0}/(\bar{e}p_{j0}+\bar{q}ep_{j1})\geq  1/2,\\
				\frac{q\bar{e}p_{j0}}{\bar{e}p_{j0}+\bar{q}ep_{j1}} , &   \text{else},
			\end{matrix}\right.\\
		\end{split}
	\end{equation}
	and
	\begin{equation}\label{d1012}
		\begin{split}
			d_{1j}=\left\{\begin{matrix}
				1-\frac{qep_{j0}}{ep_{j0}+\bar{q}\bar{e}p_{j1}},  &   qep_{j0}/(ep_{j0}+\bar{q}\bar{e}p_{j1})\geq  1/2,\\
				\frac{qep_{j0}}{ep_{j0}+\bar{q}\bar{e}p_{j1}} , &   \text{else}.
			\end{matrix}\right.
		\end{split}
	\end{equation}
	
	Applying  Corollary \ref{co2} to Example 2, we get the optimal estimator for $z=0$ and $z=1$, 
	\begin{align*}
		\hat{s}^*(x, z)= \begin{cases}0, & \text { if } x=1, \\ \operatorname{argmax}_{s \in\{0,1\}} P_S(s), & \text { if } x=0 .\end{cases}
	\end{align*}
	In fact, the estimator acquires the state knowledge completely due to $z=as+(1+N)\mod2$ whenever $x=1$. In this case, we have $c(x=1)=0$. For $x=0$, the estimator does not get any useful information about the state and we adopt the best constant estimator here, i.e., $	\hat{s}^*=\underset{s \in\{0,1\}}{\operatorname{argmax}} P_S(s)$. In this case, we have that
	\begin{align}
		c(x=0) &=\mathrm{E}\big[d\big(S, \underset{s \in\{0,1\}}{\operatorname{argmax}} P_S(s)\big) | X=0\big] \notag\\
		&=\min _{s \in\{0,1\}} P_S(s)=\min \{q, 1-q\},
	\end{align}
	where we used the independence of $S$ and $X$. Thus, the expected distortion of the optimal estimator is
	$
		D=\sum_x P_X(x) c(x)=P_X(0) c(0)=p \min \{q, 1-q\},
	$
	Therefore, we obtain 
	\begin{align*}
		C(D)&=\max_{p\in[0,1]}\{\min\{q{\rm{H}}_2(p),I(X;Z)\}:  p\min \{q, 1-q\}\le D\}.		
	\end{align*}	
	
		\subsection {Two upper bounds to Example 2}
	Applying  Theorem \ref{tho1} to Example 2 and combining the proof  in the previous subsection, we obtain 
	\begin{align*}
		C(D)&=\max_{p\in[0,1]}\{{\rm{H}}_2(p):\  p\min \{q, 1-q\}\le D\}.		
	\end{align*}

Note that the rate-distortion function for Bernoulli source with Hamming distortion is that
	\begin{align*}
		R_S(D)= \begin{cases}{\rm{H}}_2(q)-{\rm{H}}_2(D), & \text { if } 0\leq D< q, \\ 0, & \text { if } D\geq q .\end{cases}
	\end{align*}
We then apply Theorem~\ref{tho} to Example 2 and obtain
	\begin{align}
		C(D)\le 	\begin{cases} \max_{{P_{UX}}:I(U,S ; Z) \geq R_S(D)} \big\{I(X ; Y|U,S)+I(U,S ; Z)\big\}-R_S(D), & \text { if } 0\leq D< q, \\ \max_{P_{UX}} \big\{I(X ; Y|U,S)+I(U,S ; Z)\big\}, & \text { if } D\geq q .	\end{cases}
	\end{align}

	\bibliography{IEEEabrv,myref}

\end{document}